\documentclass[prd,onecolumn,nofootinbib,aps,tightenlines,preprintnumbers,superscriptaddress,11pt]{revtex4}

\usepackage{amsmath}
\usepackage{amssymb}
\usepackage{amsfonts}
\usepackage{fouridx}
\usepackage{hyperref}
\usepackage{graphicx}
\usepackage{cancel}
\usepackage[normalem]{ulem}

\def\({\left(}
\def\){\right)}
\def\[{\left[}
\def\]{\right]}
\def\llangle{\left\langle}
\def\rrangle{\right\rangle}
\def\Hilbert{\mathcal H}
\def\F{\mathcal{F}}
\def\tr{\mathop{\rm Tr}\nolimits}
\def\x{{\bf x}}

\def\k{{\bf k}}
\def\q{{\bf q}}

\newcommand{\eins}{\mbox{$1 \hspace{-1.0mm} {\bf l}$}}
\newcommand{\bra}[1]{\langle #1 |}
\newcommand{\ket}[1]{|#1\rangle}
\newcommand{\braket}[2]{\langle #1 \vphantom{#2}|
   #2 \vphantom{#1} \rangle}
\newcommand{\bracket}[3]{\langle #1 \vphantom{#2#3}|
  #2 | #3 \vphantom{#1#2} \rangle}

\newcommand{\draftnote}[1]{\textnormal{#1}}

\begin{document}

\preprint{\hbox{CALT-2016-033, UH-511-1272-2016}}

\title{How Decoherence Affects the Probability of Slow-Roll Eternal Inflation}

\author{Kimberly~K.~Boddy}
\email{kboddy@hawaii.edu}
\affiliation{Department of Physics and Astronomy, University of Hawai'i}
\author{Sean~M.~Carroll}
\email{seancarroll@gmail.com}
\author{Jason~Pollack}
\email{jpollack@caltech.edu}
\affiliation{Walter Burke Institute for Theoretical Physics, California Institute of Technology}

\begin{abstract}
Slow-roll inflation can become eternal if the quantum variance of the inflaton field around its slowly rolling classical trajectory is converted into a distribution of classical spacetimes inflating at different rates, and if the variance is large enough compared to the rate of classical rolling that the probability of an increased rate of expansion is sufficiently high.
Both of these criteria depend sensitively on whether and how perturbation modes of the inflaton interact and decohere.
Decoherence is inevitable as a result of gravitationally-sourced interactions whose strength are proportional to the slow-roll parameters.
However, the weakness of these interactions means that decoherence is typically delayed until several Hubble times after modes grow beyond the Hubble scale.
We present perturbative evidence that decoherence of long-wavelength inflaton modes indeed leads to an ensemble of classical spacetimes with differing cosmological evolutions.
We introduce the notion of per-branch observables---expectation values with respect to the different decohered branches of the wave function---and show that the evolution of modes on individual branches varies from branch to branch.
Thus single-field slow-roll inflation fulfills the quantum-mechanical criteria required for the validity of the standard picture of eternal inflation.
For a given potential, the delayed decoherence can lead to slight quantitative adjustments to the regime in which the inflaton undergoes eternal inflation.
\end{abstract}

\maketitle


\newpage
\tableofcontents

\newpage
\section{Introduction}

The state of the early universe -- hot, dense, and very smooth -- is extremely fine-tuned by conventional dynamical measures~\cite{Carroll:2014uoa}.
Inflationary cosmology~\cite{Guth:1980zm,Linde:1981mu,Albrecht:1982wi} attempts to account for this apparent fine-tuning by invoking a period of accelerated expansion in the very early universe.
The potential energy of a slowly rolling scalar field, the inflaton, serves as a source of quasi-exponential expansion through the Friedmann equation, leading to a universe that is nearly smooth and spatially flat.

Quantum mechanics, however, changes this picture of slow-roll inflation in an important way.
Although the classical equations of motion completely determine the behavior of the inflaton zero mode (i.e. the expectation value of the field) rolling down the potential, quantum field theory in curved spacetime dictates that each Fourier mode of the field has a nonzero variance (two-point function).
This variance persists after a mode leaves the Hubble radius and classically freezes out, and it is still present when inflation ends and the mode re-enters the Hubble radius.
If reheating at the end of inflation produces a sufficiently rich thermal bath of particles and radiation, decoherence~\cite{Zeh:1970fop,Zurek:1981xq,Griffiths:1984rx,Joos:1984uk,Schlosshauer:2003zy} occurs (if it has not already): the thermal bath becomes entangled with definite values of the curvature perturbation entering the Hubble radius, so that the quantum states corresponding to different values of the inflaton field become orthogonal and evolve without interference~\cite{Polarski:1995jg,Lombardo:2005iz,Martineau:2006ki,Burgess:2006jn,Burgess:2014eoa,Kiefer:2006je,Prokopec:2006fc,Liu:2016aaf}.
Hence, any modes within a Hubble volume after the end of inflation have inevitably undergone decoherence; our observable universe, including the Cosmic Microwave Background (CMB) and large-scale structure, is one branch of the universal quantum state.

Slow-roll \emph{eternal} inflation occurs when there is a period during which the quantum variance in the inflaton field is sufficiently large that the field may fluctuate upward on its potential~\cite{Linde:1982uu,Starobinsky:1982ee,Linde:1986fd,Creminelli:2008es,Martinec:2014uva}.
In regions where these upward fluctuations occur, the universe expands at a faster rate, and such regions come to dominate the physical volume of space.
If the probability of upward fluctuations is sufficiently high, the total volume of inflating space expands as a function of time, and inflation is eternal.
Although there are other mechanisms to achieve an eternally inflating universe, such as tunneling transitions which produce inflating bubbles \cite{Vilenkin:1983xq}, we concentrate on slow-roll eternal inflation and refer to it simply as eternal inflation throughout the paper.

Eternal inflation hinges on the idea that quantum fluctuations of the inflaton are true, dynamical occurrences.
However, quantum fluctuations become dynamical in unitary (Everettian, Many-Worlds) quantum mechanics only when decoherence and branching of the wave function occur~\cite{Boddy:2014eba}.
To put the slow-roll eternal inflation story on a firm foundation, it is therefore necessary to examine carefully just when inflationary modes decohere, and how that decoherence enables backreaction that can effect the value of the expansion rate in different regions.

In this paper we therefore investigate eternal inflation carefully from a quantum-mechanical perspective.
Following the approach of the recent work of Ref.~\cite{Nelson:2016kjm}, we work with the adiabatic curvature perturbation $\zeta$ and consider the lowest-order gravitationally-sourced interaction between modes of different wavelengths.
This interaction vanishes in the limit as slow-roll parameters go to zero, and therefore maintains the stability of pure de~Sitter space itself, where no decoherence should occur~\cite{Boddy:2014eba}.
It was shown in Ref.~\cite{Nelson:2016kjm} that this interaction decoheres the modes that we observe in the CMB on $\mathcal{O}(10)$ Hubble times after they cross the Hubble radius.
We consider the effects of this long-wavelength decoherence on the evolution of modes that still have short wavelengths compared to the Hubble radius at the time of decoherence, which we use as a proxy for the cosmological backreaction due to the decoherence.
We find that the standard lore in which eternal inflation occurs when quantum dispersion dominates over classical rolling down the potential is qualitatively correct, but we also show that the quantitative predictions of eternal inflation must be adjusted to incorporate the time it takes for gravitational interactions to bring about decoherence.

The remainder of this paper is structured as follows.
In Section~\ref{sec:basic} we review the standard picture of slow-roll eternal inflation and explain the basic quantum-mechanical picture behind our analysis.
In the next two sections we construct the technical machinery needed to establish the details of our picture of eternal inflation.
In Section~\ref{sec:decohere} we set up the general problem of finding the time evolution of the inflaton field and describe its solution by path-integral methods and Feynman diagrams.
We review the result of Ref.~\cite{Nelson:2016kjm} that gravitational backreaction decoheres super-Hubble adiabatic curvature modes during inflation.
In Section~\ref{sec:branching} we interpret this result in the language of wave function branching, and introduce the notion of observables within a particular branch, where the long-wavelength decohered modes have a definite classical value.
We describe the Feynman rules for computing these observables, and show in particular that the evolution of short-wavelength modes depends on the long-wavelength background, \draftnote{suggesting that different decohered branches have different cosmological histories}.
In Section~\ref{sec:eternal} we then use this machinery to study eternal inflation.
We consider the statistics of the daughter cosmologies that emerge from a single region of space as super-Hubble modes decohere and the wave function branches.
We write the probability of the effective upward evolution of the cosmological constant that heralds eternal inflation as a function of the inflationary potential.
\draftnote{The expression for the probability, as expected, largely reproduces previous results, with slight modifications as a result of correctly incorporating a potential-dependent time until decoherence.}
Finally, we discuss the broader implications of this work for the standard eternal inflation in Section~\ref{sec:discussion} and then conclude in Section~\ref{sec:conclusion}.

\section{The Basic Picture}\label{sec:basic}

To set the stage, let us consider this picture more closely.
In order to determine the global structure of a universe in which inflation has begun, it is necessary to consider modes which have left the Hubble radius and have yet to return---and indeed will possibly never return, due to the present acceleration of the universe.
If super-Hubble modes decohere in some particular basis, the quantum state of the universe as a whole can be written as a superposition of different states with definite values of the modes in that basis---``branches''---which do not interfere with one another.
In particular, some branches may have definite values of cosmological parameters, such as the Hubble constant, which differ from the values on the initial classical slow-roll trajectory.
Although the expectation values themselves will not change, individual classical patches after inflation may have values of the parameters that differ strongly from the expectation values.
Even if the parameters of a particular inflationary potential are chosen to produce a particular amplitude $\delta\rho/\rho$ for the density perturbations, for example, some of the classical cosmologies resulting from inflation on this potential will nevertheless have entirely different values.
If decoherence produces a distribution of Hubble constants around the classical value, there will be some branches of the wave function on which the Hubble constant grows rather than decreases monotonically according to the equations of motion and hence on which the end of inflation can be postponed indefinitely.
If these branches are common enough, the volume of inflating space may grow indefinitely.
There is no global spacelike hypersurface on which inflation ends, and the universe is in the regime of eternal inflation~\cite{Creminelli:2008es}.

It is therefore important to understand if eternal inflation actually occurs and under what conditions.
In the standard picture of inflation, the Hubble rate of expansion is determined by
\begin{equation}
  \frac{\dot{a}}{a} = \sqrt{\frac{V(\phi)}{3}} = H(\phi) \ ,
\end{equation}
where $8 \pi G=c=\hbar=1$, the dot notation indicates a derivative with respect to the physical time $t$, and $\phi\equiv\llangle \phi \rrangle + \delta \phi $ is the inflaton field.
Quantum fluctuations of $\delta \phi$ behave as~\cite{Vilenkin:1982wt,Linde:1982uu,Starobinsky:1982ee}
\begin{equation}
  \langle \delta\phi^2(t+\Delta t) \rangle - \langle \delta\phi^2(t) \rangle
  = \frac{H^3}{4\pi^2} \Delta t
  \label{eq:variance}
\end{equation}
over a time $\Delta t$.
According to the standard story, the quantum state of a mode collapses when it reaches the Hubble scale -- corresponding in our language to decoherence -- and each mode obtains a value given by the sum of its classical evolution plus a quantum fluctuation up its potential~\cite{Linde:1982uu,Starobinsky:1982ee}.
In the stochastic approximation, these super-Hubble modes are assumed to decohere quickly, and the evolution of the inflaton field is treated as a random walk on top of its classical slow-roll trajectory~\cite{Starobinsky:1982ee,Vilenkin:1982wt,Vilenkin:1983xq,Creminelli:2008es}.
In a Hubble time $\Delta t \sim H^{-1}$, the fluctuation in field value is $\Delta\phi \sim H/(2\pi)$.
If the size of these fluctuations are sufficiently large, inflation may persist due to the scalar field stochastically fluctuating up in its potential, countering the classical motion.
We will discuss this more extensively in Section \ref{sec:eternal} below.

The assumption of rapid decoherence does not necessarily hold in all circumstances, in which case eternal inflation must be treated appropriately in the context of quantum mechanics.
Let us therefore be a bit more explicit about the relationship between backreaction and decoherence, in a simplified toy-model context.

Consider a Hilbert space decomposed into two factors $\Hilbert = \Hilbert_L \otimes \Hilbert_S$, corresponding roughly to long-wavelength and short-wavelength modes.
Let $\{\ket{\phi_i}\}$ be a basis for $\Hilbert_L$ and $\{\ket{\omega_{a}}\}$ be a basis for $\Hilbert_S$.
We would like to illustrate the relationship between entanglement and backreaction.
Therefore consider a state of the form
\begin{equation}
  \ket{\Psi} = \alpha\ket{\phi_1}\ket{\omega_1} + \beta\ket{\phi_2}\ket{\omega_2}\ .
  \label{eq:toystate}
\end{equation}
For generic $\alpha$ and $\beta$ such a state is clearly entangled, but for $\alpha=1$, $\beta=0$ it is a product state, so this form suffices to examine both possibilities.

We would like to illustrate the (perhaps intuitive) fact that the evolution of the short-wavelength states can depend on that of the long-wavelength states with which they are entangled, but without entanglement it will simply depend on the long-wavelength state as a whole. 
In the absence of entanglement (and the decoherence that leads to it) there are no fluctuations or quantum jumps; in particular, it does not matter if the form of that state is that of a squeezed state \cite{Grishchuk:1990bj,Albrecht:1992kf}.

We therefore consider an interaction Hamiltonian that does not itself lead to decoherence; in other words, one that is a tensor product of operators on the two factors of Hilbert space, $\hat{H}^I = \hat{h}^{(L)} \otimes \hat{k}^{(S)}$.
The matrix elements of such a Hamiltonian in the $\{\ket{\phi_i}, \ket{\omega_{a}}\}$ basis take the form
\begin{equation}
  {H}^I_{iajb} = h^{(L)}_{ij} k^{(S)}_{ab}\ .
  \label{eq:toyint}
\end{equation}
Its action on the state (\ref{eq:toystate}) is
\begin{equation}
   \hat{H}^I|\Psi\rangle = \alpha \sum_{jb}\left(h^{(L)}_{1j}\ket{\phi_j}\right)\otimes \left(k^{(S)}_{1b}\ket{\omega_b}\right)
   + \beta \sum_{jb}\left(h^{(L)}_{2j}\ket{\phi_j}\right)\otimes \left(k^{(S)}_{2b}\ket{\omega_b}\right)\ .
\end{equation}
From this form it should be clear that the evolution of the short-wavelength modes depends on the branch of the wave function they are in.
In the $\alpha$ branch they evolve under the influence of the components $k^{(S)}_{1b}$, while in the $\beta$ branch they evolve under the influence of $k^{(S)}_{2b}$.
If the state were unentangled, there would be no differentiation in how different parts of the long-wavelength state might affect the evolution of the shorter modes.
In this way, decoherence is necessary for backreaction to occur differently within different branches.
It is therefore important to examine the rate of decoherence during inflation to accurately calculate the stochastic evolution of the inflationary spacetime on each branch.

\section{Gravitational Decoherence during Inflation}\label{sec:decohere}

We would like to understand the full quantum dynamics of the inflaton field during slow-roll inflation.
Following \cite{Nelson:2016kjm}, we write down an expression for the wave function and then extract information about particular modes of interest.
We confine ourselves in this section and the next to perturbative quantum field theory in curved spacetimes rather than full nonperturbative quantum gravity, so we carry out the calculations on a fixed de~Sitter background.
We argue below that our perturbative results, when appropriately interpreted, nevertheless suffice to determine how backreaction alters the effective Hubble constant and hence determine when eternal inflation occurs.
Since we are tracking the evolution of the wave function, we work in the Schr\"{o}dinger picture rather than in the interaction picture used in typical flat-space QFT calculations: we view states rather than operators as evolving in time, and our expectation values are always with respect to the wave function at the time of interest rather than $S$-matrix elements.

\subsection{The General Problem}\label{sub:general}

We want to consider the (coordinate or conformal) time evolution of (particular modes of) a quantum state $\ket\Psi$ in the Hilbert space $\Hilbert_\zeta$ of a quantum field theory of a single real scalar field $\zeta$ with translationally and rotationally invariant interactions.
A natural basis spanning this Hilbert space is the basis of field configurations, which we can think of either as functions of position space $\zeta(\x)$ or, more often, as functions of momentum space $\zeta(\k)$.
We decompose $\Hilbert_\zeta$ into an infinite tensor product of factors representing each point in (position or momentum) space,
\begin{equation}
  \Hilbert_\zeta = \bigotimes_\k \Hilbert_{\zeta,\k} \ ,
\end{equation}
so that a particular field configuration $\ket{\zeta}$ is the product of a specific multi-particle state in each individual Hilbert space factor,
\begin{equation}
  \ket{\zeta}=\bigotimes_\k \ket{\zeta_\k} \ .
\label{eq:zeta}
\end{equation}
Each $\ket{\zeta_\k}$ is an eigenstate of the field value operator $\hat \zeta_\k$ on the appropriate factor $\Hilbert_{\zeta,\k}$:
\begin{equation}
  \hat \zeta_\k \ket{\zeta_\k}=\zeta_\k \ket{\zeta_\k}\label{eq:zeta_k} \ .
\end{equation}
Thus a field configuration $\ket{\zeta}$ is a simultaneous eigenstate of all operators which consist of the tensor product of the field value operator in a given Hilbert space factor $\Hilbert_{\zeta,\k}$ and the identity in all other factors.
The collection of all of the eigenvalues $\zeta_\k$ comprises the field configuration as a function of momentum space, $\zeta(\k)$.

Given this basis, it is often convenient to work with the wave functional $\Psi[\zeta]$ instead of the state itself:
\begin{equation}
  \Psi[\zeta]\equiv\braket{\zeta}{\Psi} \ .
\end{equation}
We work in the Schr\"{o}dinger picture and consider states rather than operators as evolving in time.
Time evolution is generated by the Hamiltonian $\hat H(t)$; the symmetry assumptions mean that can we decompose it as a sum of symmetry-respecting polynomial interactions among the fields $\zeta_\k$ and the canonical momenta $\pi_\k^{(\zeta)} \equiv -i(\delta/\delta\zeta_{-\k})$.
The lowest-order terms, up to quadratic order in the fields, make up the free Hamiltonian $\hat H_\textrm{free}$.
Given $\hat H_\textrm{free}$, we can write a special Gaussian state $\ket{\Psi_G}$, which is the superposition of field configuration basis states with coefficients given by the weight $\Psi_G[\zeta](t)$ that solves the Schr\"{o}dinger equation:
\begin{equation}
  \ket{\Psi_G}=\sum_\zeta \Psi_G[\zeta] \ket{\zeta}, \hspace{0.7cm}
  i\frac{d}{dt}\Psi_G[\zeta]=\hat H_{{\rm free}}[\zeta](t)\Psi_G[\zeta] \ .
  \label{eq:sch_free}
\end{equation}
This weight is given by a Gaussian integral over the field modes:
\begin{equation}
  \Psi_G[\zeta](t) \equiv
  N_\zeta(t) \exp \[-\int_{\k} \zeta_\k \zeta^\dagger_\k A_\zeta (k,t)\] \ ,
  \label{eq:Psi_zeta}
\end{equation}
where $N_\zeta(t)$ is a normalization constant, the shorthand notation for the integral is given by Eq.~\eqref{eq:int_k} below, the complex conjugate (denoted with $\dagger$) enforces the reality condition on $\zeta(\x)$, and $A_\zeta$ depends only on the magnitude of $k$ by the symmetry assumption.
The function $A_\zeta (k,t)$ is given implicitly by Eq.~\eqref{eq:sch_free}, and we derive it explicitly for our Hamiltonian of interest below.

We assume that the initial (at $t=0$ or equivalently $\tau=-\infty$) state is simply
\begin{equation}
  \Psi[\zeta](t=0)=\Psi_G[\zeta](t=0) \ .
\end{equation}
Our assumption is motivated by the fact that this state has the form of the Euclidean\footnote{The Euclidean vacuum is also known as the Bunch-Davies vacuum \cite{Bunch:1978yq,Bunch:1978yw} for a massive, noninteracting scalar field or the Hartle-Hawking vacuum \cite{Hartle:1976tp} for an interacting one.} vacuum~\cite{Geheniau:1968bcs,Schomblond:1976xc,Chernikov:1968zm,Tagirov:1972vv,Mottola:1984ar,Allen:1985ux}, the unique state which is both de~Sitter-invariant and well-behaved at short distances, i.e.\ obeys the Hadamard condition \cite{Candelas:1975du}.
Nevertheless, it is an assumption: it implies in particular that short-wavelength modes which have just crossed the Planck scale and entered the domain of validity for QFT are in their vacuum state and unentangled with modes of different wavelengths.

\subsection{The Free Action}

We now specialize to the case of interest: perturbations around a de~Sitter background.
The background de~Sitter metric in a flat slicing is $ds^2 = -dt^2 + a(t)^2\ d\mathbf{x}^2$, where $a(t) = e^{Ht} = -1/H\tau$.
Concentrating solely on scalar modes, we work in a gauge in which fluctuations are represented as perturbations $\zeta$ of the induced spatial metric,
\begin{equation}
  g_{ij} = a(t)^2 e^{2\zeta(\mathbf{x},t)} \ .
\end{equation}
This curvature perturbation describes the amount of expansion at any point; if $\zeta \ll 1$, it describes the expansion in the given region.
The quadratic action for $\zeta$ is
\begin{equation}
  S_\textrm{free}\[\zeta\]
  = \frac{1}{2} \int d^4x \frac{2\epsilon M_p^2}{c_s^2}a^3
  \left[\dot{\zeta}^2 - \frac{c_s^2}{a^2} (\partial_i \zeta)^2 \right] \ ,
  \label{eq:S_free}
\end{equation}
where $M_p \equiv 1/\sqrt{8\pi G}$ is the reduced Planck mass and $\epsilon \equiv -\dot{H}/H^2 \ll 1$ is the first slow-roll parameter.
We set the propagation speed to $c_s=1$; Appendix B of Ref.~\cite{Nelson:2016kjm} treats the general case.
We work in Fourier space, using the conventions in Appendix~\ref{sec:A_zeta}.
Note that because $\zeta(\mathbf{x},t)$ is real we have $\zeta_\mathbf{k}^\dagger = \zeta_{-\mathbf{k}}$, at least classically.
It is also true quantum-mechanically if the quantum state is invariant under $\mathbf{k} \to -\mathbf{k}$, which is the case for our initial vacuum state.
In Appendix~\ref{sec:A_zeta} we use the free action (\ref{eq:S_free}) to derive the free Hamiltonian
\begin{equation}
  \hat H_{{\rm free}}[\zeta] = \frac{1}{2}\int_{\k}
  \left[\frac{1}{2\epsilon M_p^2 a^3}
    \pi^{(\zeta)}_\mathbf{k}\pi^{(\zeta)}_{-\mathbf{k}}
    +2\epsilon M_p^2 a k^2 \zeta_\mathbf{k}\zeta_{-\mathbf{k}}\right] \ ,
  \label{eq:H_free}
\end{equation}
and hence an expression for $A_\zeta$,
\begin{equation}
  A_\zeta(k,\tau) = k^3 \frac{\epsilon M_p^2}{H^2}
  \frac{1-\frac{i}{k\tau}}{1+k^2\tau^2} \ .
  \label{eq:green}
\end{equation}

\subsection{Interactions}\label{sub:interactions}

Thus far we have worked only with the free Hamiltonian $\hat H_{\rm free}\[\zeta\]$.
The full Hamiltonian consists of the free term and an interaction term: $\hat H[\zeta]=\hat H_{{\rm free}}[\zeta]+\hat H_{{\rm int}}[\zeta]$.
If the interaction Hamiltonian is perturbatively small, evolution with the full Hamiltonian instead of the free one adds an extra multiplicative term to the wave functional:
\begin{equation}
\Psi[\zeta](t)=\Psi_G[\zeta](t)\times\Psi_{NG}[\zeta](t).\label{eq:Psi}
\end{equation}
The lowest-order interaction is cubic, so the non-Gaussian factor can be written
\begin{equation}
  \Psi_{NG}[\zeta](t) \equiv \exp
  \[\int_{\k,\k^\prime,\q} \zeta_\k \zeta_{\k^\prime} \zeta_\q \F_{\k,\k^\prime,\q}(t)\],
\end{equation}
where the shorthand notation for the integral, which includes a momentum-conserving delta function, is given by Eq.~\eqref{eq:int_k} below.
Because we have taken $\hat H_\textrm{int}$ to be rotationally invariant, $\F_{\k,\k^\prime,\q}$ depends only the magnitudes $k$, $k^\prime$, and $q$ of the momenta.

We solve for $\F$ by writing the Schr\"{o}dinger equation using $H[\zeta]$ and then subtracting the free Schr\"{o}dinger equation.
Intuitively, $\F(\tau)$ represents the cumulative effect of all three-point interactions from the initial (conformal) time $\tau_0$ to time $\tau$.
Each specific interaction is computed by using the free Hamiltonian to evolve up to an intermediate time $\tau^\prime$, then inserting the interaction term at that time; the full effect is the result of integrating over all these intermediate times.
The result is
\begin{equation}
  \F_{\k,\k^\prime,\q}(t) = i \int_{\tau_0}^\tau \frac{d\tau^\prime}{H \tau^\prime}
  \tilde \Hilbert^{(\rm int)}_{\k,\k^\prime,\q}(\tau)
  \exp \[-i \int_{\tau^\prime}^{\tau^{\prime \prime}} d\tau^{\prime \prime}
  \alpha_{\k,\k^\prime,\q}(\tau^{\prime \prime})\] \ ,
\end{equation}
where $\tilde \Hilbert^{(\rm int)}$ is a classical source, defined implicitly through the action of $\hat H_{\rm int}$ on $\Psi_G$,
\begin{equation}
  \hat H_{\rm int}[\zeta](t) \Psi_G[\zeta](t) \equiv
  \[\int_{\k,\k^\prime,\q} \zeta_\k \zeta_{\k^\prime} \zeta_\q
  \tilde \Hilbert^{(\rm int)}_{\k,\k^\prime,\q}(t)  \] \Psi_G[\zeta](t) \ .
\end{equation}
The quantity $\alpha$ implements the free evolution,
\begin{equation}
  \alpha_{\k,\k^\prime,\q}(\tau) \equiv
  \[f_\zeta (k,\tau) A_\zeta (k,\tau)
  + f_\zeta (k^\prime,\tau) A_\zeta (k^\prime,\tau)
  + f_\zeta (q,\tau) A_\zeta (q,\tau)\] / (H \tau),
  \label{eq:alpha}
\end{equation}
where $f_\zeta$ is the coefficient of the kinetic term in $H_{{\rm free}}[\zeta]$,
\begin{equation}
  f_\zeta(k, \tau) = \frac{1}{2\epsilon M_p^2 a^3}
  = -\frac{\tau^3 H^3}{2\epsilon M_p^2} \ .
  \label{eq:fzeta}
\end{equation}
Note that $\F_{\k,\k^\prime,\q}$ is completely symmetric in its three momentum arguments.

The physically relevant interaction term for the case of interest here is the gravitationally sourced $\zeta\zeta\zeta$ interaction which contains no time derivatives and hence does not vanish in the super-Hubble limit, where $\dot\zeta$ terms are redshifted away.
We have defined $\zeta$ as the fluctuations around a de~Sitter background, so the interaction terms should vanish in the limit of pure de~Sitter space, i.e.\ they should have coefficients proportional to the slow roll parameters $\epsilon$ and $\eta$.
In particular, the interaction Hamiltonian is~\cite{Maldacena:2002vr,Nelson:2016kjm}
\begin{equation}
  \hat H_{\rm int}\[\zeta\] = \frac{M_p^2}{2} \int d^3 \x \epsilon
  \(\epsilon + \eta\) a \zeta^2 \partial^2\zeta \label{eq:3_point} \ .
\end{equation}
This expression for $\hat H_{\rm int}$ then sets the form of $\F$; the computation is performed in Ref.~\cite{Nelson:2016kjm}, which finds in particular that in the late-time limit $\tau\rightarrow 0$ the imaginary part of $\F$ dominates, $\left|{\rm Re}\, \F\right| \ll \left|{\rm Im}\, \F\right|$.
This means $\Psi_{NG}$ can be approximated as a pure phase, $\left|\Psi_{NG}[\zeta](t)\right|^2\approx1$.

\subsection{Feynman Rules}\label{sub:feynman}

In order to address the issue of backreaction, it is necessary to extend the results of Ref.~\cite{Nelson:2016kjm} by going beyond the pure-phase approximation.
Given the expression in Eq.~\eqref{eq:Psi}, we can proceed to calculate expectation values of observables.
In particular, we are interested in the evolution of short-wavelength, sub-Hubble modes.
The free evolution of a mode is given by Eq.~\eqref{eq:green}, which appears in the computation of the two-point function $\llangle \zeta_{\k_\star} \zeta_{-\k_\star} \rrangle$.

We begin by converting the operator expectation value into a path integral.
For convenience we write the path integral over field configurations as $\int \mathcal{D}\zeta \equiv \int_\zeta$.
Inserting a complete sets of states with a definite field value in each momentum mode, we have
\begin{align}
  \llangle \zeta_{\k_\star} \zeta_{-\k_\star} \rrangle
  &= \bracket{\Psi}{\hat \zeta_{\k_\star} \hat \zeta_{-\k_\star}}{\Psi} \nonumber\\
  &= \bracket{\Psi}{\(\int_{\zeta} \ket{\zeta} \bra{\zeta}\)
    \hat \zeta_{\k_\star} \hat \zeta_{-\k_\star}
    \(\int_{\zeta^\prime} \ket{\zeta^\prime} \bra{\zeta^\prime}\)}{\Psi} \nonumber\\
  &= \int_{\zeta} \Psi^\dagger[\zeta] \zeta_{\k_\star} \zeta_{-\k_\star} \Psi[\zeta]
  \label{eq:2point_Psi_squared} \\
  &= \frac{1}{N} \int_{\zeta} \zeta_{\k_\star} \zeta_{-\k_\star}
  \Psi_G^{\dagger} \Psi_G \Psi^\dagger_{NG} \Psi_{NG} \label{eq:2point} \ .
\end{align}
To lowest order in ${\rm Re}\, \F / {\rm Im}\, \F$, $\Psi_{NG}$ is a pure phase, so $\Psi^\dagger_{NG} \Psi_{NG}\approx1$ and the path integral becomes Gaussian:
\begin{align}
  \llangle \zeta_{\k_\star} \zeta_{-\k_\star} \rrangle
  &\approx \frac{1}{N} \int_{\zeta}
  \zeta_{\k_\star} \zeta_{-\k_\star} \Psi_G^{\dagger} \Psi_G \nonumber\\
  &=\frac{\int_\zeta \zeta_{\k_\star} \zeta_{-\k_\star} \exp \left\{-\int_{\k}
    \zeta_\k \zeta^\dagger_\k \[A^\dagger_\zeta (k,t) + A_\zeta (k,t) \] \right\}}
    {\int_\zeta \exp \left\{-\int_{\k} \zeta_\k \zeta^\dagger_\k
    \[A^\dagger_\zeta (k,t) + A_\zeta (k,t) \]\right\}}\label{eq:2point_fraction}\\
   &= \frac{(2\pi)^3 \delta^3(0) }{4\, \mathrm{Re}\, A_\zeta\(\k_\star,t\)}
       \label{eq:2point_tree} \ ,
\end{align}
recovering the free evolution\footnote{Our expression differs by a factor of 2 from that in Eqs.~(4.8-9) of Ref.~\cite{Nelson:2016kjm}, but as noted in Appendix~\ref{sec:A_zeta} our definition of $A_\zeta$ itself also differs by a factor of 2 and the two factors cancel here.}.
Recall again that we are working in the Schr\"{o}dinger picture, where the time dependence lives in the state $\ket{\Psi}$ rather than the operators, so the details of the calculation differ from the more familiar computation of the 2-point correlator from the path integral in QFT (though it should give the same result); in particular note that because of the $\Psi^\dagger \Psi$ term it is not the action $S$ itself but rather $S+S^\dagger=2\:{\rm Re}S$ that appears in the exponential.

We see that the pure phase assumption ensures that the (even-point) correlation functions are unchanged by the interactions.
Thus, to capture the effect of these interactions, we need to go beyond the pure phase assumption by writing the full expression for $\Psi^\dagger_{NG} \Psi_{NG}$ rather than simply approximating it as 1.
We find
\begin{align}
  \Psi^\dagger_{NG} \Psi_{NG}
  &= \exp \[\int_{\k,\k^\prime,\q} \zeta_\k \zeta_{\k^\prime} \zeta_\q
    \F_{\k,\k^\prime,\q} + \int_{\k,\k^\prime,\q}\(
    \zeta_\k \zeta_{\k^\prime} {\zeta}_\q \F_{\k,\k^\prime,\q} \)^\dagger \] \nonumber\\
  &= \exp \[\int_{\k,\k^\prime,\q} \zeta_\k \zeta_{\k^\prime} \zeta_\q
    \F_{\k,\k^\prime,\q} + \int_{\k,\k^\prime,\q} \zeta_{-\k} \zeta_{-\k^\prime}
      {\zeta}_{-\q} \F^\dagger_{\k,\k^\prime,\q} \] \nonumber\\
  &= \exp \[\int_{\k,\k^\prime,\q} \zeta_\k \zeta_{\k^\prime} \zeta_\q
    \( \F_{\k,\k^\prime,\q} + \F^\dagger_{\k,\k^\prime,\q}\)\]\label{eq:Psi_NG_squared} \\
  &= \exp \[\int_{\k,\k^\prime,\q} 2 \zeta_\k \zeta_{\k^\prime} \zeta_\q
       {\rm Re}\, \F_{\k,\k^\prime,\q} \] \label{eq:NG_squared} \ .
\end{align}
To obtain Eq.~\eqref{eq:Psi_NG_squared}, we substitute $\k,\k^\prime,\q\rightarrow-\k,-\k^\prime,-\q$ in the second integrand, which leaves the integral unchanged, keeping in mind that $\F_{\k,\k^\prime,\q}$ depends only on the magnitude of the momenta.
As desired, the imaginary part of $\F$ drops out entirely, and the integrand vanishes in the limit ${\rm Re}\, \F \rightarrow 0$.

We now insert our improved expression for $\Psi^\dagger_{NG} \Psi_{NG}$ into the two-point function $\llangle \zeta_{\k_\star} \zeta_{-\k_\star} \rrangle$ \eqref{eq:2point}:
\begin{equation}
  \llangle \zeta_{\k_\star} \zeta_{-\k_\star} \rrangle
  = \frac{1}{N} \int_{\zeta} \zeta_{\k_\star} \zeta_{-\k_\star}
  \exp \[-\int_{\k} 2 \zeta_\k \zeta^\dagger_\k \, \mathrm{Re}\, A_\zeta (k,t) \]
  \exp \[\int_{\k,\k^\prime,\q} 2 \zeta_\k \zeta_{\k^\prime} \zeta_\q \,
       {\rm Re}\, \F_{\k,\k^\prime,\q} \] \ .
\end{equation}
Since we cannot integrate this expression analytically, we Taylor-expand the interaction term, assuming that each term in the integral is perturbatively small:
\begin{equation}
  \exp \[\int_{\k,\k^\prime,\q} 2 \zeta_\k \zeta_{\k^\prime} \zeta_\q \,
  {\rm Re}\, \F_{\k,\k^\prime,\q} \]
  = 1 + \int_{\k,\k^\prime,\q} 2 \zeta_\k \zeta_{\k^\prime} \zeta_\q \,
  {\rm Re}\, \F_{\k,\k^\prime,\q}
  + \ldots
  \label{eq:taylor}
\end{equation}
We see that we can straightforwardly calculate the correlation functions using a Feynman diagram expansion, with the propagator given by $1/\[4 A_\zeta (k,t)\]$ and a single three-point interaction with coefficient $2\, {\rm Re}\, \F_{\k,\k^\prime,\q}$.

\subsection{Decoherence}

Thus far we have written down an expression \eqref{eq:Psi} for the wave functional $\Psi\[\zeta\](t)$, and hence the wave function is
\begin{equation}
  \ket{\Psi(t)}=\int_{\zeta}\Psi\[\zeta\](t)\ket{\zeta} \ .
\end{equation}
Using this expression we can compute expectation values by writing them as a path integral which admits a solution using the Feynman diagrams.

This is not, however, all that can be done with the wave function.
We have seen in the previous subsection that computing expectation values of the fields alone yields an expression (e.g.\ Eq.~\eqref{eq:2point_Psi_squared}) that depends only on the wave functional as $\Psi^\dagger \Psi$.
Such expectation values depend only on the magnitude of the the wave function, not its phase.
In addition to expectation values, we can also construct the density operator $\hat \rho\equiv\ket{\Psi}\bra{\Psi}$, which has complex matrix elements $\rho[\zeta,\zeta']=\Psi[\zeta]\Psi^\dagger[\zeta']$.
In particular, we can factorize Hilbert space by partitioning the wavenumbers, assigning those above a cutoff $\Lambda$ to the ``system'' and those below $\Lambda$ to the ``environment,''
\begin{equation}
  \ket{\zeta}=\ket{S}\ket{E}, \hspace{0.7cm}
  \Hilbert_\zeta=\Hilbert_S \otimes \Hilbert_E \ ,
\end{equation}
where
\begin{equation}
  \ket{S} = \bigotimes_{|\k|> \Lambda} \ket{\zeta_\k}, \hspace{0.7cm}
  \ket{E} = \bigotimes_{|\k|\leq\Lambda} \ket{\zeta_\k} \ .
\end{equation}
We can then write the reduced density matrix of the system
\begin{align}
  \rho_S[S,S']
  &=\bra{S}\hat{\rho}_S\ket{S'}
  =\bra{S}\tr_E\(\ket{\Psi}\bra{\Psi}\)\ket{S'} \nonumber\\
  &=\bra{S}\int\mathcal{D}E\; \braket{E}{\Psi}\braket{\Psi}{E}\ket{S'}
  =\int\mathcal{D}E\; \Psi[S,E]\Psi^\dagger[S',E]
\end{align}
where in the last step we have defined the wave functional $\Psi[S,E]$ as the matrix element between $\ket{S}\ket{E}$ and $\ket{\Psi}$:
\begin{equation}
  \Psi[S,E]=\(\bra{S}\otimes\bra{E}\)\ket{\Psi} \ .
\end{equation}

Decoherence occurs in the system when interactions between the system and the environment cause the decoherence factor (the ratio of the off-diagonal elements of $\rho_S$ to the diagonal ones) to become small:
\begin{equation}
  D[S,S'] \equiv \frac{\left|\rho_S[S,S']\right|}
  {\sqrt{\rho_S[S,S]\rho_S[S',S']}}\ll 1 \ .
\end{equation}
Inserting our expression for $\Psi$ \eqref{eq:Psi} and noting that the Gaussian part \eqref{eq:Psi_zeta} factors as $\Psi_G[\zeta]= \Psi^{(S)}_G[S](t) \times \Psi^{(E)}_G[E](t)$, the decoherence factor becomes
\begin{align}
  D[S,S']
  &=\left|\frac{\int\mathcal{D}E\; \Psi[S,E]\Psi^\dagger[S',E]}
  {\sqrt{\(\int\mathcal{D}E\; \Psi[S,E]\Psi^\dagger[S,E]\)
      \(\int\mathcal{D}E\; \Psi[S',E]\Psi^\dagger[S',E]\)}} \right| \nonumber\\
  &= \left|\frac{\int\mathcal{D}E\; \left|\Psi^{(E)}_G[E]\right|^2
    \Psi_{NG}[S,E] \Psi^\dagger_{NG}[S',E]}
  {\sqrt{\(\int\mathcal{D}E\; \left|\Psi^{(E)}_G[E]\right|^2
  \left|\Psi_{NG}[S,E] \right|^2\)
  \(\int\mathcal{D}E\; \left|\Psi^{(E)}_G[E]\right|^2
  \left|\Psi_{NG}[S',E]\right|^2\)}} \right| \ .
\end{align}
When the non-Gaussian piece of the wave function is a pure phase, which is the case to lowest order in ${\rm Re}\, \F/{\rm Im}\, \F$ in Section~\ref{sub:interactions}, both integrals in the denominator integrate to one and the decoherence factor simplifies to
\begin{equation}
  D[S,S'] = \left|\int\mathcal{D}E\; \left|\Psi^{(E)}_G[E]\right|^2
  \Psi_{NG}[S,E] \Psi^\dagger_{NG}[S',E] \right|\label{eq:simple_D} \ .
\end{equation}

The problem is now reduced to performing the calculation with the previously given forms of $\Psi_G$ and $\Psi_{NG}$.
Ref.\ \cite{Nelson:2016kjm} carries out this calculation for the case of a single super-Hubble mode, $\Hilbert_S=\{\zeta_\q,\zeta^\dagger_\q=\zeta_{-\q},\:q<H\}$.
As in Section~\ref{sub:feynman}, Eq.~\eqref{eq:simple_D} can be written as an expectation value, this time in the theory of the environment modes, and solved in the deeply super-Hubble limit $|q\tau|\ll 1$ using Feynman diagrams and the cumulant expansion.
In our notation, the result is \cite{Nelson:2016kjm}
\begin{equation}
D[\zeta_\q,\tilde{\zeta}_\q](\tau=-1/aH) = \exp\[-\frac{1}{288}(\epsilon+\eta)^2|\Delta\bar{\zeta}_\q|^2 \(\frac{aH}{q}\)^3+\ldots\],\
\end{equation}
where the dots indicate terms higher-order in $\F^2$ and $\Delta\bar{\zeta}_\q\equiv \bar \zeta - \bar{\zeta}^\prime_\q$ is the rescaled dimensionless amplitude of $\zeta_\q-\zeta^\prime_\q$, defined by $\zeta_\q\equiv V^{1/2} q^{-3/2} \pi \sqrt{2} \bar{\zeta}_\q$.
The barred quantities have variance
\begin{equation}
  \llangle|\bar \zeta_\q|^2\rrangle
  =\frac{H^2}{2 \epsilon M_p^2}\frac{1}{(2 \pi)^2} \equiv \Delta_{\zeta}^2 \ ,
  \label{eq:delta_zeta}
\end{equation}
and so $\llangle|\Delta\bar{\zeta}_\q|^2\rrangle=2\Delta_{\zeta}^2$.
The dimensionless decoherence ``rate'' is then the negative log of the decoherence factor with $|\Delta\bar{\zeta}_\q|^2$ set equal to its expectation value,
\begin{equation}
  \Gamma_{\rm deco} (q,a) \approx
  \(\frac{\epsilon+\eta}{12}\)^2 \Delta_\zeta^2 \(\frac{aH}{q}\)^3
  \hspace{0.7cm}\textrm{for}\hspace{0.25cm} q\ll aH \ .
  \label{eq:gamma}
\end{equation}
Decoherence has occurred when this rate, and hence the negative of the exponent in the decoherence factor, becomes large.
The rate does not grow large until long after Hubble crossing, at $q=aH$, because of the smallness of the slow-roll parameters and the amplitude of fluctuations (constrained by observations of the CMB \cite{Ade:2015xua} to be $\Delta_\zeta^2\sim10^{-9}$ at 60 $e$-folds before the end of inflation).
For reasonable values of $(\epsilon + \eta)\sim10^{-5}$--$10^{-2}$, the modes seen in the CMB would have decohered 10--20 $e$-folds after Hubble crossing.

In the remainder of this paper, we discuss the implications of this delayed decoherence for eternal inflation.
In the next section, we establish that decoherence of long-wavelength modes affects the evolution of short-wavelength modes evolving in the decohered long-wavelength background, and argue that this change in evolution implies the backreaction of the Hubble constant required for eternal inflation.
We then turn to discussion of the quantitative differences between the resulting picture and the standard picture of stochastic eternal inflation caused by the delay of decoherence far beyond Hubble crossing.

\section{Branching and Backreaction}\label{sec:branching}

As we have shown, the results of Ref.~\cite{Nelson:2016kjm} indicate that decoherence of super-Hubble modes due to gravitational interactions alone is inevitable, though the weakness of these interactions means that the modes typically take several Hubble times after Hubble crossing to decohere.
Because modes continually expand across the Hubble radius during inflation, they are also continually decohering, so the overall wave function is itself continually branching; on each branch there is a definite classical value for every mode which has become sufficiently long-wavelength.
Since long-wavelength and short-wavelength modes interact gravitationally, we expect the short-wavelength modes to have a different reaction in different branches to the decohered long-wavelength modes.

In this section we formalize this argument, which we have already made schematically in Section \ref{sec:basic}, by introducing the notion of per-branch observables.
We show that short-wavelength modes do indeed evolve differently in different branches of the wave function.
We argue that this differing evolution indicates the nature of backreaction away from our perturbative picture; short-wavelength modes evolve as if they experience different values of the Hubble constant in different branches, and there exists a gauge choice on which the effective Hubble constant itself differs from branch to branch.

\subsection{Observables on Branches}

In the previous section, we calculated the expectation value of products of fields with respect to the overall state $\ket{\Psi}$.
Once decoherence has occurred, however, the evolution in a particular decohered branch is not given by this expectation value, but from the expectation value with respect to the state of that particular branch.
As discussed in Section \ref{sub:general} above, every field configuration $\ket{\zeta}$ is an eigenstate of field value for each individual momentum mode.
Since the mode decoheres in the field value basis, we can label individual branches by the field value of the decohered mode\footnote{In fact modes larger than $\k_{\rm dec}$ have also decohered, so properly speaking we must specify the values of all the decohered modes to uniquely label a branch. We neglect this complication, which can easily be incorporated at the cost of complicating the notation, throughout the section. The final Feynman rules presented in Fig.~\ref{fig:feynman}, however, take the need to consider each decohered mode into account.} in that branch, $\zeta_{\k_{\rm dec}}^\star$.
The state $\ket{\Psi}$ can thus be projected onto an individual branch by considering only the field configurations on which the field value of the decohered mode is $\zeta_{\k_{\rm dec}}^\star$, then renormalizing.

More precisely, we define the state $\ket{\zeta_{\k}^\star}\in \Hilbert_{\zeta,\k}$ as the eigenstate of $\hat \zeta_{\k}$ with eigenvalue $\zeta_{\k}^\star$, as in Eq.~\eqref{eq:zeta_k}.
Then $\ket{\zeta_{\k}^\star}\bra{\zeta_{\k}^\star}$ projects states in the Hilbert space factor $\Hilbert_{\zeta,\k}$, and we can define an associated projector on the entire Hilbert space $\Hilbert_\zeta$ by multiplying this projector by the identity on all other factors,
\begin{equation}
  \hat P^{\zeta_{\k}^\star}_\k \equiv
  \(\ket{\zeta_{\k}^\star}\otimes\bigotimes_{\k^\prime\ne\k}\eins_{\k^\prime}\)
  \(\bra{\zeta_{\k}^\star}\otimes\bigotimes_{\k^\prime\ne\k}\eins_{\k^\prime}\) \ ,
\end{equation}
whose action on field configurations defined by Eq.~\eqref{eq:zeta} is simply
\begin{equation}
  \hat P^{\zeta_{\k}^\star}_\k \ket{\zeta}
  = \braket{\zeta_{\k}^\star}{\zeta_\k} \ket{\zeta}
  = \delta_{\zeta_{\k}^\star,\zeta_\k}\ket{\zeta} \ .
\end{equation}

We can now repeat the calculation in Section~\ref{sub:feynman} above for a branch with a definite field value $\zeta_{\k_{\rm dec}}^\star$ in the $\k_{\rm dec}$-th mode:
\begin{align}
  \llangle \zeta_{\k_\star} \zeta_{-\k_\star} \rrangle_{\zeta_{\k_{\rm dec}}^\star}
  &= \frac{1}{N_{\zeta_{\k_{\rm dec}}^\star}} \bracket{\Psi}
     {\hat P^{\zeta_{\k_{\rm dec}}^\star}_{\k_{\rm dec}} \hat \zeta_{\k_\star}\hat \zeta_{-\k_\star}
     \hat P^{\zeta_{\k_{\rm dec}}^\star}_{\k_{\rm dec}}}{\Psi} \nonumber\\
  &= \frac{1}{N_{\zeta_{\k_{\rm dec}}^\star}}\bracket{\Psi}
     {\(\int_{\zeta} \ket{\zeta} \bra{\zeta}\)\hat
     P^{\zeta_{\k_{\rm dec}}^\star}_{\k_{\rm dec}}\hat \zeta_{\k_\star}\hat \zeta_{-\k_\star}\hat
     P^{\zeta_{\k_{\rm dec}}^\star}_{\k_{\rm dec}}
     \(\int_{\zeta^\prime} \ket{\zeta^\prime} \bra{\zeta^\prime}\)}{\Psi} \nonumber\\
  &= \frac{1}{N_{\zeta_{\k_{\rm dec}}^\star}}\int_{\zeta} \Psi^\dagger[\zeta]
     \int_{\zeta^\prime} \bra{\zeta}\hat P^{\zeta_{\k_{\rm dec}}^\star}_{\k_{\rm dec}}
     \hat \zeta_{\k_\star}\hat \zeta_{-\k_\star} \hat P^{\zeta_{\k_{\rm dec}}^\star}_{\k_{\rm dec}}
     \ket{\zeta^\prime}\Psi[\zeta^\prime] \nonumber\\
  &= \frac{1}{N_{\zeta_{\k_{\rm dec}}^\star}}\int_{\zeta} \Psi^\dagger[\zeta]
     \int_{\zeta^\prime} \zeta_{\k_\star} \zeta^\prime_{-\k_\star}
     \braket{\zeta_{\k_{\rm dec}}}{\zeta_{\k_{\rm dec}}^\star}
     \braket{\zeta_{\k_{\rm dec}}^\star}{\zeta^\prime_{\k_{\rm dec}}}
     \braket{\zeta}{\zeta^\prime}\Psi[\zeta^\prime] \nonumber\\
  &= \frac{1}{N_{\zeta_{\k_{\rm dec}}^\star}}\int_{\zeta}
     \braket{\zeta_{\k_{\rm dec}}^\star}{\zeta_{\k_{\rm dec}}}^2
     \Psi^\dagger[\zeta] \zeta_{\k_\star} \zeta_{-\k_\star} \Psi[\zeta] \nonumber\\
  &= \frac{1}{N_{\zeta_{\k_{\rm dec}}^\star}} \int_{\zeta}
     \braket{\zeta_{\k_{\rm dec}}^\star}{\zeta_{\k_{\rm dec}}}^2 \zeta_{\k_\star}
     \zeta_{-\k_\star} \Psi_G^{\star} \Psi_G \Psi^\dagger_{NG} \Psi_{NG} \ ,
  \label{eq:2point_dec}
\end{align}
where the normalization factor is defined so that the wave function on each branch has unit norm, $\bracket{\Psi}{\hat P^{\zeta_{\k_{\rm dec}}^\star}_{\k_{\rm dec}}\hat P^{\zeta_{\k_{\rm dec}}^\star}_{\k_{\rm dec}}}{\Psi}/N_{\zeta_{\k_{\rm dec}}^\star}=1$.
Again, Eq.~\eqref{eq:2point_dec} says that we are supposed to integrate only over the field configurations where the decohered mode has the correct field value, i.e.\ the ones on the appropriate branch.

\subsection{Feynman Rules on Branches}

In the pure-phase approximation, the integrals over $\zeta_{\k_{\rm dec}}$ and $\zeta_{\k_\star}$ are independent and the extra term contributes only an overall constant of proportionality that cancels in the normalization.
In this approximation, evolution of short-wavelength modes is unaffected by decoherence.
A better approximation is to treat ${\rm Re}\, \F$ as small compared to ${\rm Im}\, \F$, yielding Eq.~\eqref{eq:NG_squared}.
Inserting this quantity into the two-point function for a decohered branch $\llangle \zeta_{\k_\star} \zeta_{-\k_\star} \rrangle_{\zeta_{\k_{\rm dec}}^\star}$, Eq.~\eqref{eq:2point_dec} gives
\begin{align}
  \llangle \zeta_{\k_\star} \zeta_{-\k_\star} \rrangle_{\zeta_{\k_{\rm dec}}^\star}
  = \frac{1}{N_{\zeta_{\k_{\rm dec}}^\star}} \int_{\zeta}
  \braket{\zeta_{\k_{\rm dec}}^\star}{\zeta_{\k_{\rm dec}}}^2
  \zeta_{\k_\star} \zeta_{-\k_\star}
  & \exp \[-\int_{\k} 2 \zeta_\k \zeta^\dagger_\k \mathrm{Re}\, A_\zeta (k,t) \]
  \nonumber\\
  & \times \exp \[\int_{\k,\k^\prime,\q} 2 \zeta_\k \zeta_{\k^\prime} \zeta_\q
       {\rm Re}\, \F_{\k,\k^\prime,\q} \] \ . \label{eq:2point_dec_subsitituted}
\end{align}
This expression, combined with the Taylor expansion (\ref{eq:taylor}), allows us to compute correlation functions on decohered branches, but actually writing down the equivalent Feynman rules requires some thought.
Ultimately (from the path-integral perspective) we can use Feynman diagrams to compute correlation functions because Taylor expansion lets us write each integral over momentum modes in the form of a polynomial multiplied by a Gaussian in a particular momentum mode, which we can compute using Wick's theorem.
Only the integrals for which the polynomial is a nontrivial function of the momentum modes yield nontrivial results; the contribution of every other Gaussian is canceled by the denominator.
In terms of Feynman diagrams, these canceled expressions are just the disconnected diagrams.
For example, in computing the propagator in Eq.~\eqref{eq:2point_tree} from Eq.~\eqref{eq:2point_fraction}, only the terms in the exponential with $\k=\pm \k_\star$ are important.

We can use Feynman diagrams to compute correlation functions in a particular branch, but we need to carefully take into account the extra factor of $\braket{\zeta_{\k_{\rm dec}}^\star}{\zeta_{\k_{\rm dec}}}^2$, i.e.\ we need to restrict the path integral to only span over field configurations with nonzero overlap with the branch.
This gives a delta function for each decohered mode.
We could impose the delta function separately on each diagram containing decohered modes, but we may also immediately use the delta function to integrate over these modes and simplify the path integral.
We integrate each integral over the decohered field mode $\zeta_{\k_{\rm dec}}$ by localizing to the actual value of the mode on the branch, replacing $\zeta_{\k_{\rm dec}}$ by $\zeta_{\k_{\rm dec}}^\star$ wherever it appears.

One replacement is in the $\zeta_{\k_{\rm dec}}^2$ term that is the coefficient of $\mathrm{Re}\, A_\zeta (k_{\rm dec},t)$ in (\ref{eq:2point_dec_subsitituted}).
After we have made this replacement, this term yields a $\zeta$-independent normalization factor which cancels in the numerator and denominator.
In terms of Feynman diagrams, the propagator factor for a decohered momentum mode is just $1$, which is unsurprising because we have set this mode equal to its classical value in the branch.
At this point we can simply integrate out the propagating decohered modes entirely; all interactions involving them will involve the insertion of a classical external source.

\begin{figure}[t]
  \begin{center}
    \includegraphics[width=0.9\textwidth]{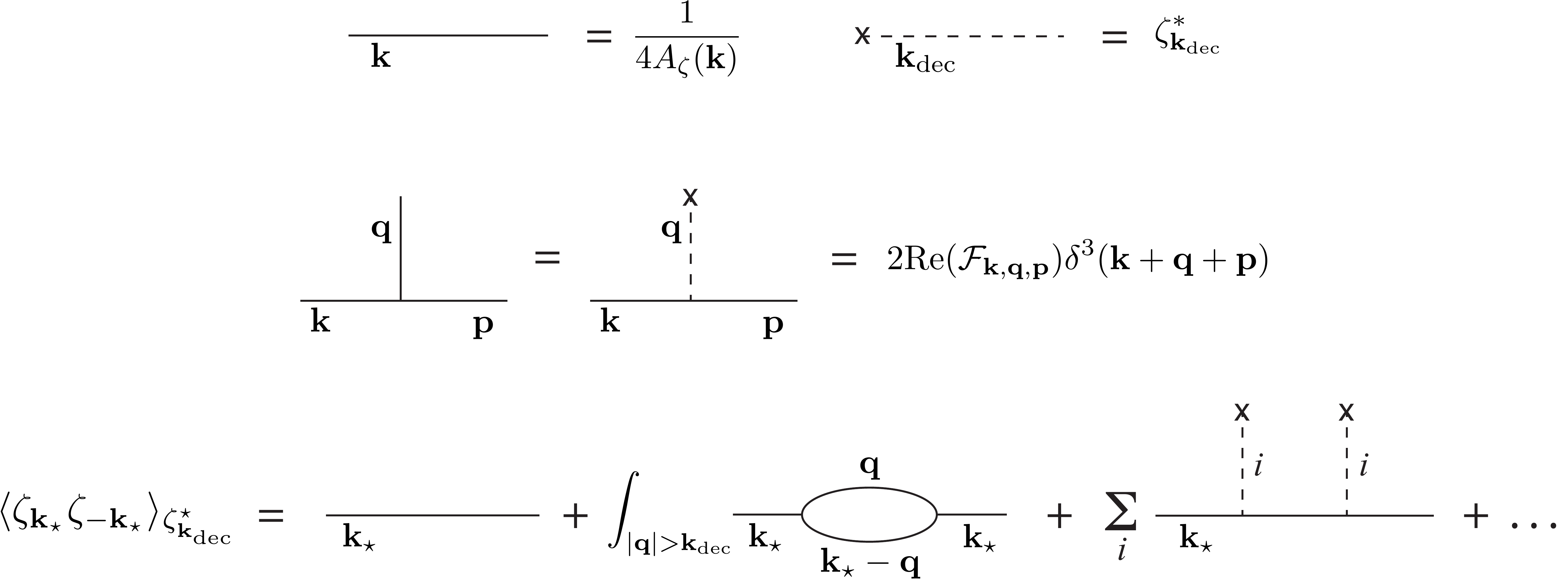}
  \end{center}
  \caption[Feynman diagrams]{Computation of $\llangle \zeta_{\k_\star} \zeta_{-\k_\star} \rrangle_{\zeta_{\k_{\rm dec}}^\star}$ using Feynman diagrams.
As discussed in Section \ref{sub:feynman} above, non-decohered modes have a propagator given by $1/\[4 A_\zeta (k,t)\]$ and a single three-point interaction with coefficient $2\, {\rm Re}\, \F_{\k,\k^\prime,\q}$.
A decohered mode field insertion comes with a factor of its field value, $\zeta_{\k_{\rm dec}}^\star$.
At leading order the $\zeta_{\k_\star}$ two-point function is corrected by diagrams with two interaction vertices.
We split the diagrams into two categories: those where no intermediate momenta are decohered, which we write as a loop correction integrating over momenta greater than $\k_{\rm dec}$, and those involving decohered momenta, which we represent as a sum over diagrams with two field insertions.}
\label{fig:feynman}
\end{figure}

In addition, we need to replace the decohered field modes which appear in the interaction term.
We treat each such mode as a frozen classical source, to be inserted as necessary in the propagator for the dynamical short-wavelength modes, as shown in Fig.~\ref{fig:feynman}.
Our assumption of perturbativity allows us to approximate the interaction term by its Taylor expansion truncated at a given order, yielding a polynomial in $\zeta_{\k}$.
The delta function means that we need to replace the polynomial with a piecewise function which substitutes $\zeta_{\k_{\rm dec}}^\star$ for $\zeta_{\k_{\rm dec}}$ on configurations that overlap with the branch and is zero on all other configurations.
Again, this substitution takes place in both the numerator and the denominator (normalization factor).
At the level of the first quantum corrections, only the lowest-order term in the denominator (the zero-interaction term, with no factors of $\zeta_{\k_{\rm dec}}^\star$) contributes, so there is a contribution to the path integral with two insertions of the decohered modes\footnote{Because interactions conserve momentum, the term with one insertion does not contribute to $\llangle \zeta_{\k_\star} \zeta_{-\k_\star} \rrangle_{\zeta_{\k_{\rm dec}}^\star}$, which has equal ingoing and outgoing short-wavelength momentum.}, proportional to $(\zeta_{\k_{\rm dec}}^{\star})^2$.
In terms of Feynman diagrams, each insertion of an external decohered mode gives a factor of $\zeta_{\k_{\rm dec}}^\star$.
As expected, the leading correction to the two-point function of a non-decohered field is proportional to the square of the field value of the classical $\zeta_{\k_{\rm dec}}$ field.
This confirms our intuition that short-wavelength modes should evolve differently in different branches.

In summary, the Feynman rules, shown in Fig.~\ref{fig:feynman}, are the following.
For non-decohered fields, the propagator is $1/\(4 A_\zeta (k,t)\)$. 
For each decohered field $\zeta_{\k_{{\rm dec},i}}$ labeled by $i$, only modes with the specific decohered field value $\zeta_{\k_{{\rm dec},i}}^\star$ contribute on a given branch, and only as external sources.
For these modes, field insertions give a factor of $\zeta_{\k_{{\rm dec},i}}^\star$. 
All three-point functions among decohered and non-decohered fields have the same interaction vertex, with coefficient $2 {\rm Re}\, \F_{\k,\k^\prime,\q}$.

\subsection{Cosmological Evolution}\label{sec:evolution}

In the previous subsection we established the intuitive result that short-wavelength modes evolving in a particular branch are affected by long-wavelength modes as if they are evolving in a particular classical background\footnote{In single-field slow-roll inflation, the three-point function $\langle \zeta_\q \zeta_{\k_\textrm{ph}} \zeta_{\k^\prime_\textrm{ph}}\rangle^\prime$ in ``physical coordinates'' vanishes in the squeezed limit, $\q \rightarrow 0$~\cite{Pajer:2013ana,Tanaka:2011aj}, where $k_\textrm{ph}\equiv k(1-\zeta_L)$ and the prime indicates the removal of the momentum-conserving delta function.
The vanishing correlation between short-wavelength modes and long-wavelength modes in these coordinates might seem in contradiction with our claim that the evolution of the short-wavelength modes depends on the value of the long-wavelength modes.
However, decoherence does not change the value of expectation values with respect to the overall wave function $\ket{\Psi}$.
Our claim is that the evolution of short-wavelength modes \textit{on each individual branch} depends on the long-wavelength field values which characterize the branch.
As previously discussed, this evolution is distinct from the evolution of short-wavelength modes in the overall wave function.
The short-wavelength modes are thus uncorrelated with long-wavelength modes in expectation values with respect to the overall wave function, but not with respect to individual branches.}, namely the solution to the Einstein equations with the particular nonzero values of the $\zeta$ field at long wavelengths (i.e.\ field values $\zeta_{\k_{\rm dec}}^\star$) that characterize the branch.
In general these geometries, unlike our initial background cosmology, will have nonzero (and nontrivial) spatial curvature.
Reproducing the usual eternal inflation story requires transforming to a gauge where the spatial curvature is once again zero, in which we expect that the geometries on various branches of the wave function will have different Hubble constants.
This is a standard procedure in the eternal inflation literature (see e.g.\ Ref.~\cite{Baumann:2009ds}) and we only sketch out the steps schematically.

We first switch from the $\zeta$ basis, where the probability distribution over field values is given in the pure phase approximation by Eq.~\eqref{eq:Psi_zeta}, to the basis of inflaton field values $\phi_\k$ in which the eternal inflation picture is usually developed.
In the inflaton field gauge, the propagating degree of freedom is the variation $\delta \phi$ of the inflaton field from its expectation value.
The power spectrum is that of a light scalar field in de~Sitter space:
\begin{equation}
\llangle \delta \phi_\k \delta \phi_{\k^\prime}\rrangle = \(2 \pi \)^3 \delta\(\k + \k^\prime \) \frac{2 \pi^2}{k^3} \(\frac{H}{2 \pi}\)^2.
\end{equation}
Just as the $\zeta$ power spectrum defines the coefficient $A_\zeta(k,t)$ of the kinetic term in the action via Eq.~\eqref{eq:2point_tree}---and hence the wave function through Eq.~\eqref{eq:Psi_zeta}---the $\delta \phi$ power spectrum defines a new coefficient $A_{\delta \phi}$.
We can therefore rewrite Eq.~\eqref{eq:Psi_zeta} in the inflaton field value basis by replacing $A_\zeta(k,t) \rightarrow A_{\delta \phi} (k, t)$.
This is simply a change of variables which does not alter the wave function itself: we are merely shifting a constant factor $1/2\epsilon$ between the coefficient $A$ and the field variable.
In particular, the branching structure of the wave function itself is preserved: decoherence gives definite values of long-wavelength $\delta \phi$ modes just as it gives definite values of long-wavelength $\zeta$ modes.
For the rest of the paper, it is convenient to work with the resulting distribution of inflaton field values.

On each branch of the wave function, we treat the decohered mode as a delta-function momentum-space perturbation of the inflaton field away from its background value.
This perturbation breaks the isotropy of the system, so we can no longer solve for the cosmological evolution using the Friedmann equations, but we can instead use perturbation theory around the initial de~Sitter background (e.g.\ Ref.~\cite{Dodelson:2003ft}) to compute the shift in the spatial geometry.
Finally, we change gauges to one in which the spatial part of the metric is again homogenous and isotropic.
This yields a probability distribution over de~Sitter regions with different values of the Hubble parameter $H$, producing branches on which inflation proceeds at different rates.
The usual practice in the eternal inflation literature is to instead say that inflation proceeds at different rates in separate spatial regions in a single overall spacetime.
We will comment further on this interpretation in the Discussion below.

\section{Eternal Inflation}\label{sec:eternal}

Our goal in this section is to consider how the classical picture of slow-roll inflation, in which the cosmology of a region of space undergoing inflation simply responds to the expectation value of the inflaton field, is modified when we include decoherence and branching.
Following the existing literature on eternal inflation and the stochastic approximation, we work directly with Fourier modes of the inflaton field $\phi$ rather than the adiabatic curvature perturbation $\zeta$.
As noted in the previous section, even though we established decoherence in the $\zeta$ field value basis, branches with definite values of $\zeta_\k$ should also have definite values of $\phi_\k$.

\subsection{The Distribution of Branches after Decoherence}\label{sub:branch_distribution}

Although we have seen that modes are continually decohering as they grow larger than the decoherence scale $\k_{\rm dec}^{-1}$, it suffices to follow the evolution of one particular mode, with expectation value $\phi_\star$ at the time it grows beyond the Hubble radius.
First consider the classical evolution.
Recall the Friedmann equations:
\begin{equation}
  H^2=\rho/3, \hspace{0.7cm} \frac{\ddot{a}}{a}=-\(\rho+3 p\)/6 \ ,
\end{equation}
where as in Section \ref{sec:basic} we have set $8\pi G = c = 1$.
A scalar field obeys the Klein-Gordon equation,
\begin{equation}
  \ddot{\phi}+3 H \dot{\phi} = -V^\prime \ ,
\end{equation}
where $^\prime=d/d\phi$, and has energy density
\begin{equation}
  \rho=\dot{\phi}^2/2+V(\phi) \ .
\end{equation}
In the slow-roll regime, $\ddot \phi \ll 3 H \dot \phi, -V^\prime$ and $\dot \phi^2 \ll V$, and the field value evolves classically at a rate
\begin{equation}
  \dot \phi=-\frac{V^\prime}{3 H} \ .
\end{equation}
In one Hubble time the classical change is therefore
\begin{equation}
  \Delta\phi_{\mathrm{c}}\equiv \dot\phi H^{-1} = -\frac{V^\prime}{3 H^2} \ .
  \label{eq:classical_change}
\end{equation}
Meanwhile, the dispersion around the classical value~\cite{Vilenkin:1982wt,Vilenkin:1983xq,Linde:1982uu,Starobinsky:1982ee,Creminelli:2008es} obeys Eq.~\eqref{eq:variance}, so the variance accumulated in a single Hubble time is
\begin{equation}
  \Delta_{\mathrm{q}}^2 \equiv \left(\langle {\delta\phi}^2(t+\Delta t) \rangle
  - \langle {\delta\phi}^2(t) \rangle\right)_{\Delta t = H^{-1}}
  = \frac{H^2}{4\pi^2} \ .
\end{equation}
\draftnote{The overall variance of $\delta\phi$ continues to grow as modes expand past Hubble crossing, but the variance of individual modes freezes out once they exceed the Hubble scale, with variance $\Delta_{\mathrm{q}}^2$.}

We are interested in what happens after $N$ $e$-folds after Hubble crossing, where $N$ is the number of $e$-folds at which modes decohere, which we write explicitly for a general slow-roll potential $V(\phi)$ below.
At this time the particular mode we are following, now with size $\lambda_{\rm dec}\equiv e^N H^{-1}$, decoheres into branches.
On each branch of the wave function, the mode has a definite classical value, and the probability distribution of these classical values is given by a Gaussian with width $\Delta_{\mathrm{q}}$ and mean $\phi_\star+N\Delta\phi_{\mathrm{c}}$:
\begin{equation}
  P(\phi)\equiv\frac{1}{\sqrt{2\pi \Delta_{\mathrm{q}}^2}}
  \exp{\[-\frac{\(\phi-\phi_\star-N\Delta\phi_{\mathrm{c}}\)^2}
    {2\Delta_{\mathrm{q}}^2} \]} \ ,
  \label{eq:prob_phi}
\end{equation}
where the prefactor ensures proper normalization of the probability distribution.

Note that $V^\prime$ and $H$ are both properly functions of $\phi$, so the classical change $\Delta\phi_{\mathrm{c}}$ also depends on the inflaton's location on the potential.
In Eq.~\eqref{eq:prob_phi} we have neglected this effect and assumed that $\Delta\phi_{\mathrm{c}}$ is constant over the range of field values we are interested in, so that the total classical rolling over $N$ $e$-folds is just $N\Delta\phi_{\mathrm{c}}$. 
We will relax this assumption below when we consider corrections to the standard eternal inflation picture.

\subsection{The Regime of Eternal Inflation}

\begin{figure}[t]
  \begin{center}
    \includegraphics[width=0.4\textwidth]{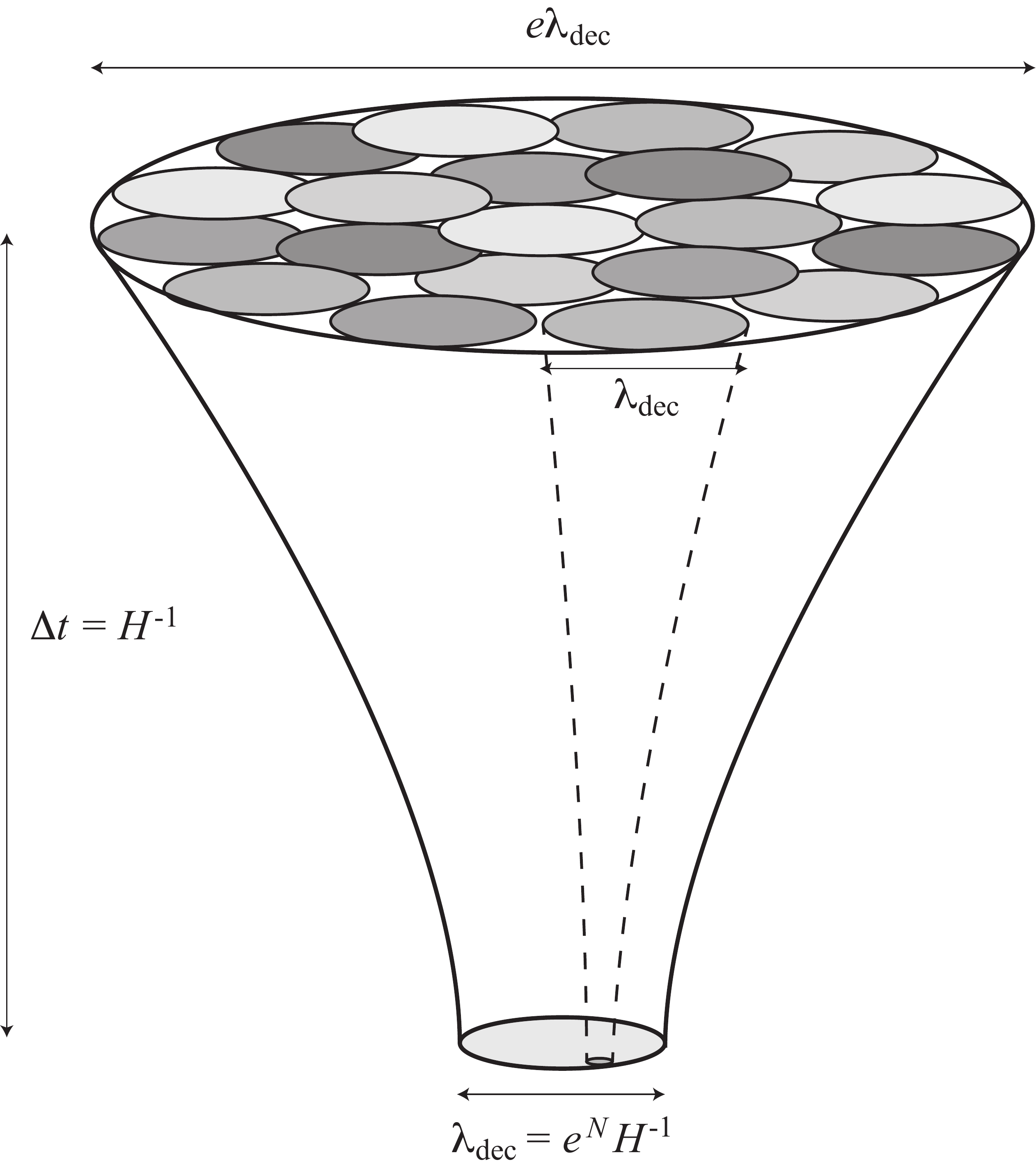}
  \end{center}
  \caption[Eternal inflation]{The evolution of patches in eternal inflation. We choose to look at an initial patch of linear size given by the wavelength at which modes decohere, $N$ $e$-folds after Hubble crossing, $\lambda_\mathrm{dec} = e^N H^{-1}$. One Hubble time later, the linear size of this comoving region has expanded by $e$, so the volume now contains $e^3 \approx 20$ patches of the size of the original region.}
\label{fig:eternal_diagram}
\end{figure}

Eq.~\eqref{eq:prob_phi} gives the probability distribution over field values for decohered inflaton modes.
Given this probability distribution, when does eternal inflation occur?
We are concerned with computing the change in eternal inflation due to delayed decoherence, so we first give the conventional account of eternal inflation~\cite{Linde:1982uu,Starobinsky:1982ee,Linde:1986fd,Creminelli:2008es,Martinec:2014uva}.
We need to compare the expectation value $\llangle\phi(t=t_0)\rrangle$ of the mode of interest at some initial time $t_0$ before decoherence has occurred to its value in particular decohered branches, drawn from the probability distribution $P(\phi)$, which is defined at the time of decoherence, $t=t_0+\Delta t$.
The probability that the field on a particular branch has moved up its potential is given by
\begin{equation}
  \textrm{Pr}(\phi>\llangle\phi(t=t_0)\rrangle)
  \equiv\int_{\llangle\phi(t=t_0)\rrangle}^\infty P(\phi) d\phi \ .
\end{equation}
Because $P(\phi)$ is supported on all values of $\phi$, the probability that the field on a particular branch has moved up its potential is always strictly nonzero.
When the probability is large enough, however, we say that the entire ensemble of branches, i.e.\ the wave function, is undergoing eternal inflation.
Here ``large enough'' is usually taken to mean larger than the reciprocal of the growth in volume during this time: $\textrm{Pr}(\phi>\llangle\phi_0\rrangle) \gtrsim e^{-3 H \Delta t}$.

This criterion for eternal inflation to occur is usually justified in terms of the growth of the volume of inflating spacetime.
The situation is depicted in Fig.~\ref{fig:eternal_diagram}.
Consider a volume of space with initial size given by the decoherence length $\lambda_\mathrm{dec}\equiv e^N H^{-1}$.
In the time $\Delta t$ it takes for a given mode to reach the scale $\lambda_\mathrm{dec}$ and decoheres, the initial volume will have grown by a factor $e^{3 H \Delta t}$.
We can therefore divide the volume into $e^{3 H \Delta t}$ regions with volume equivalent to the initial one.
We imagine for now that decoherence results in a separate classical field value in each of these regions (we will discuss the validity of this assumption later).
Hence if the probability of moving up the potential in a given region is larger than $e^{-3 H \Delta t}$, a typical branch of the wave function describing the evolution of the entire initial volume will contain at least one region of the same size as that initial volume where the field has moved up on the potential and the rate of expansion has increased.
In this case inflation is said to be ``self-reproducing'' or eternal.
It remains only to choose a convenient timescale.
The physically relevant timescale in the problem is the Hubble time $H^{-1}$, which leads to the familiar criterion that eternal inflation occurs if there is a probability to move up the potential of at least $e^{-3}\approx5\%$.

Accordingly, consider the situation one Hubble time before decoherence occurs.
\draftnote{Subject to the assumptions discussed at the end of Subsection \ref{sub:branch_distribution},} the expectation value of the \draftnote{mode of interest} is then
\begin{equation}
  \llangle\phi(t=t_0)\rrangle=\phi_\star+(N-1)\Delta\phi_{\mathrm{c}}
  =\phi_\star + (N-1) V^\prime/3 H^2 \ ,
\end{equation}
where again $\phi_\star$ is the field value at Hubble crossing, \draftnote{while the variance, which has been frozen out since Hubble crossing, remains} $\Delta_{\mathrm{q}}^2=H^2/4 \pi^2$.
Now wait for one last Hubble time.
The volume of the inflating space expands by a factor of $e^3\approx 20$, and the expectation value of the field changes to $\phi_\star+N\Delta\phi_{\mathrm{c}}$.

The probability that the field has effectively ``jumped'' up the potential compared to where it was an $e$-fold ago is given by the proportion of the probability distribution where $\phi>\phi_\star+(N-1)\Delta\phi_{\mathrm{c}}$:
\begin{equation}
  \textrm{Pr}\(\phi>\phi_\star+(N-1)\Delta\phi_{\mathrm{c}}\)
  \equiv \int_{\phi_\star+(N-1)\Delta\phi_{\mathrm{c}}}^\infty P(\phi) d\phi
  =\frac{1}{2}\[1-\mathrm{erf}
  \(\frac{-\Delta\phi_{\mathrm{c}}}{\Delta_{\mathrm{q}} \sqrt{2}}\)\] \ .
  \label{eq:Pr}
\end{equation}
Recall that the error function erf$(x)$ ranges from $0$ to $1$ as $x$ ranges from 0 to $\infty$.
So a large probability of jumping up the potential requires that the quantum dispersion is large compared to the classical rolling.

\draftnote{Notice that the final expression in Eq.~\eqref{eq:Pr} lacks any direct dependence on $N$, the number of $e$-folds from Hubble crossing to decoherence.
Hence when the expression is valid we recover exactly the standard predictions of eternal inflation.}

We can now insert the details of the inflationary potential.
First, the argument of the error function is
\begin{equation}
  \frac{-\Delta\phi_{\mathrm{c}}}{\Delta_{\mathrm{q}} \sqrt{2}}
  =\frac{\pi \sqrt{2} V^\prime}{3 H^3 }
  =\frac{2 \pi  \sqrt{\epsilon}}{ H } \ ,
  \label{eq:cl_to_qu}
\end{equation}
where we have used $\epsilon=(V^\prime/V)^2/2$, $H^2=V/3$.
Slow-roll eternal inflation in the sense we have described above occurs when
\begin{equation}
  \textrm{Pr}\[\phi>\phi_\star+(N-1)\Delta\phi_{\mathrm{c}}\] > e^{-3} \ .
\end{equation}
Eqs.~\eqref{eq:Pr} and \eqref{eq:cl_to_qu}
\ let us check where this is true for a given potential given the Hubble parameter $H$ and slow-roll parameters $\epsilon$ and $\eta$.
We see from Eq.~\eqref{eq:cl_to_qu} that quantum fluctuations become more important for flatter potentials (small $\epsilon$) and at greater energy scales (large $H/M_p$).

\subsection{Corrections from Delayed Decoherence}

In deriving Eq.~\eqref{eq:Pr} we assumed, as discussed at the end of Subsection \ref{sub:branch_distribution}, that the rate of classical rolling $\Delta\phi_{\mathrm{c}}$ was constant over the range of $e$-folds from Hubble crossing to decoherence and hence that the total classical rolling in this time was just $N \Delta\phi_{\mathrm{c}}$.
In this subsection we investigate the slight corrections which result from relaxing this assumption.
We focus on determining the range of $\phi$ values in which modes that cross the Hubble scale freeze out with sufficiently large variance to allow for eternal inflation.

As explained in the last subsection, we are interested in the last $e$-fold of classical expansion before decoherence occurs. 
Denote the value of $\phi$ at the start of this interval by $\phi_s$ and at the end by $\phi_e$. 
As above, the value of $\phi$ when the mode of interest crossed the Hubble scale is denoted by $\phi_\star$.
We can now rewrite the probability distribution of classical field values at decoherence as
\begin{equation}
  P(\phi)\equiv\frac{1}{\sqrt{2\pi \Delta_{\mathrm{q}}^2\(\phi_\star\)}}
  \exp{\[-\frac{\(\phi-\phi_e\)^2}
    {2\Delta_{\mathrm{q}}^2} \]}
  \label{eq:prob_phi_2}
\end{equation}
and the probability of moving upward on the potential as
\begin{equation}
  \textrm{Pr}\(\phi>\phi_1\)
  \equiv \int_{\phi_s}^\infty P(\phi) d\phi
  =\frac{1}{2}\[1-\mathrm{erf}
  \(\frac{-\(\phi_s-\phi_e\)}{\Delta_{\mathrm{q}}\(\phi_\star\) \sqrt{2}}\)\] \ .
  \label{eq:Pr_2}.
\end{equation}

If the field is still in the slow-roll regime at the time that the mode of interest decoheres, Eq.~\eqref{eq:classical_change} 
is still valid:
\begin{equation}
\phi_s-\phi_e\approx\dot\phi H^{-1} = -\frac{V^\prime}{3 H^2},
\label{eq:s_minus_e}
\end{equation}
but now we should evaluate $V^\prime$ and $H$ during the last $e$-fold of inflation before decoherence,say at $\(\phi_s+\phi_e\)/2$, rather than at Hubble crossing.

We would like to evaluate Eq.~\eqref{eq:s_minus_e} and thus Eq.~\eqref{eq:Pr_2} as a function of the field value at horizon crossing, $\phi_\star$.
A first approximation is to take
\begin{equation}
\phi_s  - \phi_e\approx-\left.\frac{V^\prime}{3 H^2}\right|_{\phi=\phi_\star},
\end{equation}
but this simply reproduces the $N$-independent expression for $\textrm{Pr}(\phi)$ given in the previous expression.
If we are far enough in the slow-roll regime, $N \ddot \phi \ll 3 H \dot \phi$, we can do better by evaluating $H$ and $\phi$ at the first-order approximation to $\(\phi_s+\phi_e\)/2$, i.e.\ $\phi_\star+(N-1/2)\Delta\phi_{\mathrm{c}}$:
\begin{equation}
\phi_s  - \phi_e\approx-\left.\frac{V^\prime}{3 H^2}\right|_{\phi=\phi_\star-\(N-\frac{1}{2}\)\left.\frac{V^\prime}{3 H^2}\right|_{\phi=\phi_\star}}.
\label{eq:s_minus_e_2}
\end{equation}
This expression may then straightforwardly be evaluated for a given potential. 
Notably, a dependence on $N$ has now been reintroduced.
Using Eqs.~\eqref{eq:gamma} and \eqref{eq:delta_zeta},
\begin{equation}
  N\equiv\(\ln\frac{aH}{q}\quad\mathrm{s.t.}\ \Gamma_{\rm deco}=1\)
  =-\frac{1}{3}\ln\frac{H^2 \(\epsilon+\eta\)^2 }{1152 \pi^2 \epsilon}
  \approx 3.11 -\frac{1}{3}\ln\frac{H^2 \(\epsilon+\eta\)^2 }{\epsilon} \ .
  \label{eq:N_dec}
\end{equation}
At the order we are working it is consistent to evaluate this expression at $\phi=\phi_\star$.

\begin{figure}[t]
  \begin{center}
    \includegraphics[width=0.92\textwidth]{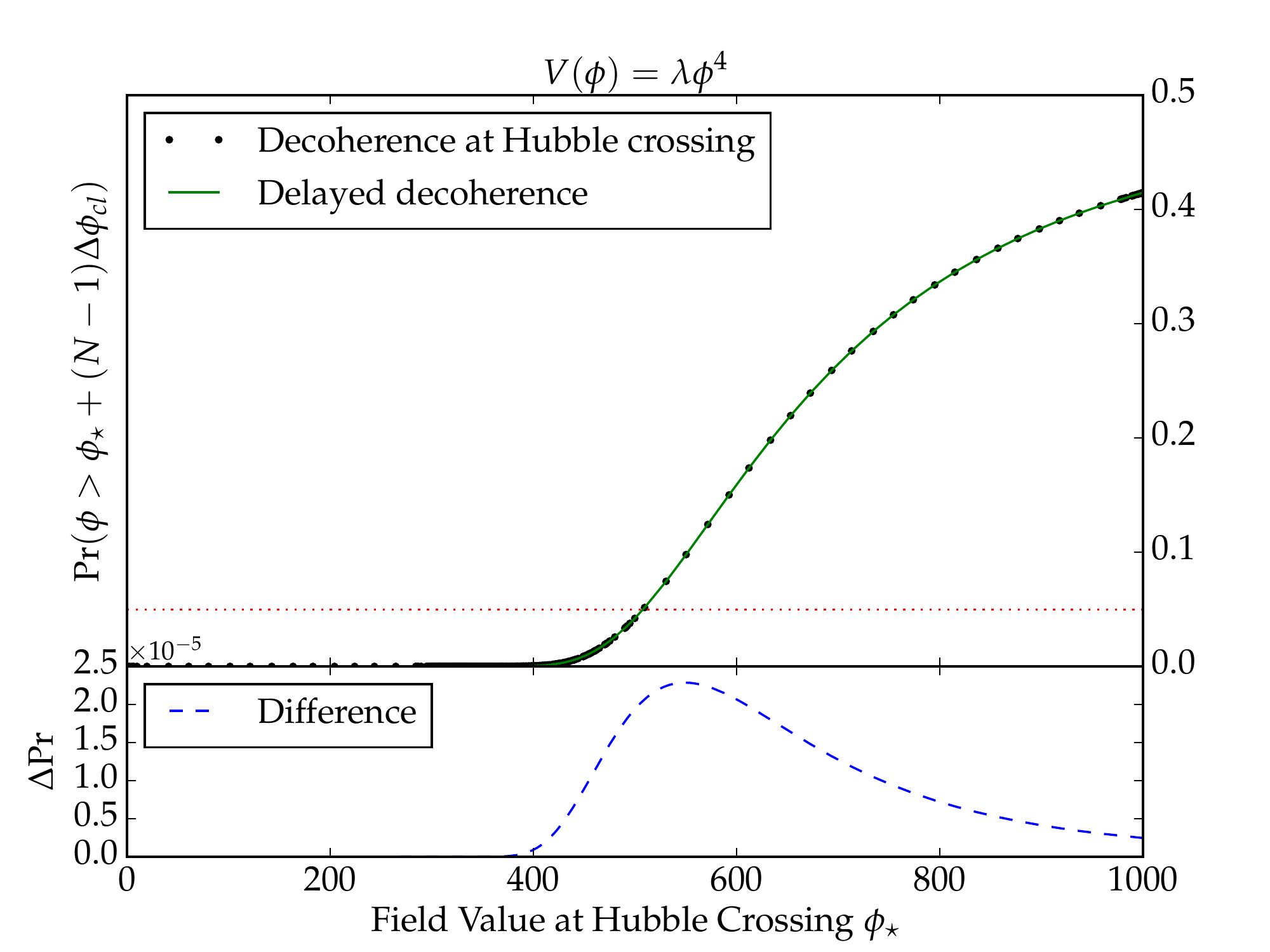}
  \end{center}
  \caption[Eternal inflation for a $\phi^4$ potential]{
    Eternal inflation for a $\phi^4$ potential.
    We have set $\lambda\approx 4.28\times10^{-14}$, which is the value required to reproduce the amplitude of fluctuations in the CMB: $\Delta_\zeta^2\approx 2.5\times10^{-9}$ 60 $e$-folds before the end of inflation.
    On the top plot, the green solid line plots the probability of eternal inflation for modes passing the Hubble scale at a field value $\phi_\star$ using Eqs.~\eqref{eq:Pr_2} and \eqref{eq:s_minus_e_2}; the black dots show the result using Eq.~\eqref{eq:Pr}.
    The red dotted horizontal line shows the probability value required for eternal inflation, $e^{-3}\approx0.05$.
    The bottom plot shows the difference between the two expressions: the difference in probabilities has a value of around $10^{-5}$ at field values $\phi_\star\sim500$ near the lower end of the regime where eternal inflation is allowed.
    The difference in probabilities is always positive because $\lambda \phi^4$ is concave up, so moving downward on the potential decreases $V^\prime$ and thus the classical rolling per $e$-fold.}
    
  \label{fig:phi4}
\end{figure}

As a worked example, Figure \ref{fig:phi4} plots the two expressions \eqref{eq:Pr} and \eqref{eq:Pr_2} for a $\phi^4$ potential. 
For this potential $N(\phi)$ decreases logarithmically with $\phi$, from 9.38 at $\phi=100$ to 7.85 at $\phi=1000$.
This delayed decoherence has only a small effect on the probability of eternal inflation, changing the probability by order $10^{-5}$.

\section{Discussion}\label{sec:discussion}

In the previous section we have largely worked within the standard picture of eternal inflation, altering it only by changing when the onset of decoherence occurs.
In the process we have noted a few uncertainties regarding this picture, which to our knowledge have not been fully resolved. 

One ambiguity is the value of $\Delta t$, the time interval at which we calculate how the wave function has branched (or in conventional language, at which quantum jumps occur).
Equivalently, this is the time before decoherence at which we take the expectation value $\llangle\phi\rrangle$, in order to compare it to the distribution $P(\phi)$ of values of the field in decohered branches, and therefore evaluate the probability that the field has jumped up in its potential, allowing for eternal inflation.
We have chosen $\Delta t=H^{-1}$, which reproduces the criterion that inflation is eternal when at least 5\% of patches have jumped upward on the potential.
Note that this implies that $N=1$ in the standard picture, which corresponds to decoherence occurring one $e$-fold after Hubble crossing, \textit{not} at Hubble crossing itself---a fact which does not seem to be commonly appreciated but is implicit in early work on eternal inflation such as Ref.~\cite{Vilenkin:1983xq}.
The criterion for when eternal inflation occurs depends on $\Delta t$, \draftnote{though only slightly, since it changes the field value at which we should evaluate the classical rolling.}

We are therefore left with the perhaps disquieting fact that whether or not inflation is eternal does not seem to be entirely objective, but rather depends on our choice of discretization.
For now, we note that two alternate choices of $\Delta t$ seem unsatisfactory.
Comparing the situation at decoherence to the situation at Hubble crossing itself, $\Delta t=N H^{-1}$, neglects the fact that in this time many other modes have decohered, making eternal inflation seem harder to achieve than it should actually be.
On the other hand, making the approximation that decoherence is instantaneous, $\Delta t=0$, in addition to being physically unrealistic, simply gives a probability of 50\% that the field value increased, which does not seem to match our intuition that eternal inflation should depend on the details of the inflaton potential.
So for the moment our choice of $\Delta t=H^{-1}$ seems most natural, in addition to most directly allowing for comparison to the standard picture.
We hope to return to this issue in future work.
One possibility is that, instead of assuming that decoherence happens immediately, we should be more careful in computing the timescale over which decoherence occurs and inserting this timescale in our calculations.
Another possibility, as we now discuss, is that the comparison of field values before and after decoherence is not the appropriate way to determine whether inflation is eternal.

A second, perhaps more serious, issue is the tension between a traditional semiclassical spacetime picture, in which branches of the wave function represent particular spacetimes in which the inflaton takes on slightly different values in nearby patches of space, versus a more intrinsically quantum picture, in which the wave function itself is primary and spacetime is emergent.
Establishing that decoherence has occurred means that we can write the wave function in terms of non-interfering branches, each of which have a definite classical value of the decohered mode.
It is not clear how we should take into account different probabilities for our universe to emerge from reheating in each of these branches (though one of us has considered a more general version of this question \cite{Sebens:2014iwa}), and/or whether we should consider the different rates of expansion in the different branches.
This question seems intimately related to the inflationary measure problem (for reviews, see, e.g., \cite{Winitzki:2006rn,Freivogel:2011eg}).
Some authors have argued that there is a coherent picture of different inflating regions as present in a single spacetime~\cite{Bousso:2011up}, others that the multiverse must be thought of as inherently quantum~\cite{Nomura:2012zb}.
We hope to consider this question more extensively in future work.
One step in this direction might include more fully carrying out the program sketched in Section~\ref{sec:evolution} to explicitly derive the wave function of an inflating scalar field in terms of branches with definite values of the Hubble parameter.

\section{Conclusion}\label{sec:conclusion}

In this paper we have tried to place the assumptions of decoherence and backreaction required for slow-roll eternal inflation on a firmer quantum-mechanical footing.
In single-field slow-roll inflation, we can definitively establish the decoherence properties of the inflaton by considering spatial perturbations around a background de~Sitter metric.
In this gauge the leading interaction is a gravitationally sourced cubic one \eqref{eq:3_point} whose strength depends on the parameters of the inflaton potential, so that in the slow-roll regime inflaton modes do not typically decohere until they have become very long-wavelength, several $e$-folds after they pass the Hubble scale \eqref{eq:N_dec}.
When decoherence has occurred, we have shown that the evolution of inflaton modes is different on different decohered branches of the wave function, each representing a different classical spacetime.
Hence the daughter cosmologies after decoherence has occurred have the differing cosmological evolutions required for the eternal inflation mechanism.
We can use this backreaction to reproduce the standard predictions for the regime of eternal inflation given a potential, and compute the (typically small) numerical changes to the boundaries of this regime.

\section*{Acknowledgements}
We thank the anonymous reviewer of the first draft of our manuscript for pointing out an error in our interpretation of Eq.~\eqref{eq:variance} which affected our numerical results.
K.B. is funded in part by DOE grant DE-SC0010504.
S.C. and J.P. are funded in part by the Walter Burke Institute for Theoretical Physics at Caltech, by DOE grant DE-SC0011632, by the Foundational Questions Institute, and by the Gordon and Betty Moore Foundation through Grant 776 to the Caltech Moore Center for Theoretical Cosmology and Physics.

\appendix
\section{Free Hamiltonian and Green Function}\label{sec:A_zeta}

In this Appendix we derive the free Hamiltonian in Eq.~\eqref{eq:H_free} and the Green function in Eq.~\eqref{eq:green} in the Schr\"{o}dinger picture.
We begin with the quadratic action for $\zeta$ \eqref{eq:S_free}, setting $c_s=1$.
To first order\footnote{It suffices to work at lowest order because the terms generated by quadratic corrections cancel in the Hamiltonian density up to cubic order; see footnote 18 of Ref.~\cite{Nelson:2016kjm}.}, the conjugate momentum of $\zeta$ is
\begin{equation}
  \pi^{(\zeta)} = \frac{\partial\mathcal{L}}{\partial\dot{\zeta}}
  = 2\epsilon M_p^2 a^3 \dot{\zeta} \ ,
\end{equation}
which obeys the canonical commutation relation $[\zeta(\mathbf{x}),\pi^{(\zeta)}(\mathbf{y})] \equiv i\delta^3(\mathbf{x}-\mathbf{y})$.
Although we will write quantities as function of $\tau$, recall that we defined the overdot notation to denote derivatives with respect to $t$.
We use the Fourier transform $\zeta_\mathbf{k} = \int d^3\mathbf{x} \zeta(\mathbf{x}) e^{-i\mathbf{k}\cdot\mathbf{x}}$ to write the conjugate momentum in terms of its wavelength modes
\begin{equation}
  \pi^{(\zeta)}_\mathbf{k} = 2\epsilon M_p^2 a^3 \dot{\zeta}_\mathbf{k} \ ,
\end{equation}
which are still functions of time.
Hence the free Hamiltonian is
\begin{align}
  \hat{H}_{\rm free}\[\zeta\] &= \int d^3\mathbf{x}
    \left[\pi^{(\zeta)}\dot{\zeta}-\mathcal{L}\right]
    = (2\epsilon M_p^2 a^3) \int d^3\mathbf{x} \left[\dot{\zeta}^2
      -\frac{1}{2}\left(\dot{\zeta}^2
      -\frac{1}{a^2}(\partial_i\zeta)^2\right)\right] \nonumber\\
    &= \frac{1}{2}\int_\k \left[\frac{1}{2\epsilon M_p^2 a^3}
      \pi^{(\zeta)}_\mathbf{k}\pi^{(\zeta)}_{-\mathbf{k}}
      +2\epsilon M_p^2 a k^2 \zeta_\mathbf{k}\zeta_{-\mathbf{k}}\right] \ ,
\end{align}
which matches Eq.~\eqref{eq:H_free}.
For convenience, we define
\begin{equation}
  \int_\k \equiv \int \frac{d^3\k}{(2\pi)^3} \qquad\textrm{and}\qquad
  \int_{\k,\k^\prime,\q} \equiv \int \frac{d^3 \k}{\(2 \pi\)^3}
  \frac{d^3 \k^\prime}{\(2 \pi\)^3}\frac{d^3 \q}{\(2 \pi\)^3}\(2 \pi\)^3
  \delta^3\(\k+\k^\prime+\q\) \ .
  \label{eq:int_k}
\end{equation}

With this Hamiltonian and the assumed form of the wave function in Eq.~\eqref{eq:Psi_zeta}, we expand both sides of the free Schr\"{o}dinger equation (\ref{eq:sch_free})
\begin{equation}
  i\frac{d}{dt}\Psi_G[\zeta](\tau)
  = \hat H_{{\rm free}}[\zeta]\Psi_G[\zeta](\tau) \ .
\end{equation}
For the left-hand side of this equation, we find
\begin{equation}
  i \frac{d}{dt} \Psi_G^{(\zeta)}[\zeta] (\tau)
  = i\Psi_G^{(\zeta)}[\zeta] (\tau) \left(\frac{\dot{N}_\zeta}{N_\zeta}
  - \int_\k \zeta_\mathbf{k} \zeta_{-\mathbf{k}} \dot{A}_\zeta(k,\tau) \right) \ .
\end{equation}
For the right-hand side, we must act with the conjugate momentum on the wave function, and thus we express it as a functional derivative: $\pi_\mathbf{k}^{(\zeta)} = -i \delta/\delta\zeta_{-\mathbf{k}}$.
We find
\begin{align}
  \pi_\mathbf{k}^{(\zeta)}\Psi_G^{(\zeta)}[\zeta](\tau)
  &= i\zeta_\mathbf{k} \left[A_\zeta(-k,\tau) + A_\zeta(k,\tau) \right]
    \Psi_G^{(\zeta)}[\zeta](\tau) \\
  \pi_{-\mathbf{k}}^{(\zeta)}\Psi_G^{(\zeta)}[\zeta](\tau)
  &= i\zeta_{-\mathbf{k}} \left[A_\zeta(-k,\tau) + A_\zeta(k,\tau) \right]
    \Psi_G^{(\zeta)}[\zeta](\tau) \\
  \pi_\mathbf{k}^{(\zeta)} \pi_{-\mathbf{k}}^{(\zeta)} \Psi_G^{(\zeta)}[\zeta](\tau)
  &= (2\pi)^3 \left[A_\zeta(-k,\tau) + A_\zeta(k,\tau) \right]
    \Psi_G^{(\zeta)}[\zeta](\tau) \nonumber\\
  &  \quad {}- \zeta_\mathbf{k} \zeta_{-\mathbf{k}}
    \left[A_\zeta(-k,\tau) + A_\zeta(k,\tau) \right]^2
    \Psi_G^{(\zeta)}[\zeta](\tau).
\end{align}
The right-hand side of the free Schr\"{o}dinger equation becomes
\begin{align}
  \hat{H}_{\rm free}(t) \Psi_G^{(\zeta)}[\zeta] (\tau)
  = \frac{1}{2} \int_\k
    & \left[(2\pi)^3 f_\zeta 2A(k,\tau) \phantom{\frac{k}{a}}\right.\nonumber\\
    & \left. - f_\zeta (2A(k,\tau))^2 \zeta_\mathbf{k}\zeta_{-\mathbf{k}}
    + \frac{1}{f_\zeta} \frac{k^2}{a^2} \zeta_\mathbf{k}\zeta_{-\mathbf{k}} \right]
     \Psi_G^{(\zeta)}[\zeta] (\tau) \ ,
\end{align}
where
\begin{equation}
  f_\zeta(\tau) \equiv \frac{1}{2\epsilon M_p^2 a^3}
  = -\frac{\tau^3 H^3}{2\epsilon M_p^2} \ .
\end{equation}
We are interested in solving for $A$, so we match the terms proportional to $\zeta_\mathbf{k}\zeta_{-\mathbf{k}}$ to obtain the differential equation
\begin{equation}
  \dot{A} = -2if_\zeta A^2 + \frac{i}{2f_\zeta}\frac{k^2}{a^2} \ .
\end{equation}
After making a change of variables to $a=\exp(Ht)$ and defining
\begin{equation}
  A = \frac{aH}{2if_\zeta(a)} \frac{du}{da}\frac{1}{u} \ ,
\end{equation}
the differential equation becomes~\cite{Burgess:2014eoa}
\begin{equation}
  a^2 \frac{d^2u}{da^2} + 4a \frac{du}{da} + \frac{k^2}{H^2 a^2} u = 0 \ ,
\end{equation}
This is the Klein-Gordon equation in de~Sitter, which can be solved in terms of Bessel functions.
We define $u = x^{3/2} y$ and change variables to $x=k/aH=-k\tau$ to obtain
\begin{equation}
  x^2 \frac{d^2y}{dx^2} + x\frac{dy}{dx} + \left(x^2-\nu^2\right)y = 0 \ ,
\end{equation}
where $\nu = 3/2$, and the solutions are the Bessel functions of the first and second kinds.
To find the correct form of $y(x)$, we apply initial condition in the far past ($a\to 0$ or $x\to\infty$ or $\tau\to -\infty$ or $t\to -\infty$) that space is de~Sitter and thus the solution is quasistatic: $dA/dt=0$.
The limiting form of $y$ becomes
\begin{equation}
  y \to u_0 x^{-3/2} e^{-ix} \ .
\end{equation}
The appropriate combination of Bessel functions that give the $\exp(-ix)$ dependence is the Hankel function of the 2nd kind, $H_\nu^{(2)}(x)$.
For $\nu=3/2$,
\begin{equation}
  y(x) = H_{3/2}(x)
  = -\sqrt{\frac{2}{\pi x}} \left(1-\frac{i}{x}\right) e^{-ix} \ .
\end{equation}
Substituting $y$ for $A$, we find
\begin{equation}
  A_\zeta(k,\tau) = k^3 \frac{\epsilon M_p^2}{H^2}
  \frac{1-\frac{i}{k\tau}}{1+k^2\tau^2} \ ,
\end{equation}
which is our desired result.
Note that this expression differs by a factor of 2 from Eq.~(5.4) of Ref.~\cite{Nelson:2016kjm}.

\bibliography{inflation-bib}
\bibliographystyle{utphys}
\end{document}